\documentclass[aps,prd,preprint,superscriptaddress,tightenlines,nofootinbib]{revtex4}

\pdfoutput=1

\usepackage{epsf}
\usepackage{dcolumn}
\usepackage{bm}
\usepackage{ifthen}

\def\epjC#1,#2(#3){{\rm Eur.\ Phys.\ J.\ }{\bf C#1}, {\rm#2} {\rm(#3)}}

\usepackage{graphicx}
\usepackage{amssymb}
\usepackage{amsfonts}
\usepackage{amsmath}
\usepackage{color}
\usepackage{ifthen}
\usepackage{rotating}
\usepackage{setspace}

\usepackage{indentfirst}

\begin{document}
\newboolean{cbxNote}
\setboolean{cbxNote}{false}
%
%
\ifthenelse{\boolean{cbxNote}}{
\preprint{CBX xx-xx}
}{
\preprint{CLNS 06/1983}
\preprint{CLEO 06-23}
}

\title{A Study of Exclusive Charmless Semileptonic $B$ Decays and \\ Extraction of $|V_{ub}|$ at CLEO}

\ifthenelse{\boolean{cbxNote}}{
\author{R. Gray}
\author{M. Shepherd}
\author{L. Gibbons}
}{
\author{D.~M.~Asner}
\author{K.~W.~Edwards}
\affiliation{Carleton University, Ottawa, Ontario, Canada K1S 5B6}
\author{R.~A.~Briere}
\author{T.~Ferguson}
\author{G.~Tatishvili}
\author{H.~Vogel}
\author{M.~E.~Watkins}
\affiliation{Carnegie Mellon University, Pittsburgh, Pennsylvania 15213}
\author{J.~L.~Rosner}
\affiliation{Enrico Fermi Institute, University of
Chicago, Chicago, Illinois 60637}
\author{N.~E.~Adam}
\author{J.~P.~Alexander}
\author{D.~G.~Cassel}
\author{J.~E.~Duboscq}
\author{R.~Ehrlich}
\author{L.~Fields}
\author{L.~Gibbons}
\author{R.~Gray}
\author{S.~W.~Gray}
\author{D.~L.~Hartill}
\author{B.~K.~Heltsley}
\author{D.~Hertz}
\author{C.~D.~Jones}
\author{J.~Kandaswamy}
\author{D.~L.~Kreinick}
\author{V.~E.~Kuznetsov}
\author{H.~Mahlke-Kr\"uger}
\author{T.~O.~Meyer}
\author{P.~U.~E.~Onyisi}
\author{J.~R.~Patterson}
\author{D.~Peterson}
\author{J.~Pivarski}
\author{D.~Riley}
\author{A.~Ryd}
\author{A.~J.~Sadoff}
\author{H.~Schwarthoff}
\author{X.~Shi}
\author{S.~Stroiney}
\author{W.~M.~Sun}
\author{T.~Wilksen}
\author{M.~Weinberger}
\affiliation{Cornell University, Ithaca, New York 14853}
% ...
\author{(CLEO Collaboration)} 
\noaffiliation
}

\date{March 23, 2007}

\begin{abstract} 

We have studied semileptonic $B$ decay to the exclusive charmless  states 
$\pi$, $\rho/\omega$, $\eta$ and $\eta^\prime$  using the full 15.5 fb$^{-1}$
CLEO $\Upsilon(4S)$  sample, with measurements performed in 
subregions of phase space to minimize dependence on {\it a priori} knowledge of
the form factors involved.  
We find total branching fractions
 ${\cal B}(B^0 \to \pi^-\ell^+\nu)=(1.37 \pm 0.15_{\text{stat}} \pm 0.11_{\text{sys}} ) \times 10^{-4}$
and ${\cal B}(B^0 \to \rho^- \ell^+\nu)=(2.93 \pm 0.37_{\text{stat}} \pm 0.37_{\text{sys}})\times 10^{-4}.$
 We find evidence  for $B^+\to\eta^{\prime}\ell^+\nu$, with 
${\cal B}(B^+\to\eta^{\prime}\ell^+\nu)=(2.66\pm \pm0.80_{\text{stat}}\pm0.56_{\text{sys}})\times 10^{-4}$
and
$1.20\times 10^{-4}<{\cal B}(B^+\to\eta^{\prime}\ell^+\nu)<4.46\times 10^{-4}$ (90\% CL).
We also limit ${\cal B}(B^+\to\eta\ell^+\nu)<1.01\times 10^{-4}$ (90\% CL). 
By combining our  $B\to\pi\ell\nu$ information with unquenched lattice 
calculations, we find  $|V_{ub}| = (3.6\pm 0.4_{\text{stat}} \pm 0.2_{\text{sys}} {}^{+0.6}_{-0.4\text{thy}})\times 10^{-3}$.

\end{abstract}

\ifthenelse{\boolean{cbxNote}}{}{
}
\maketitle

%
%

\section{Introduction}

$V_{ub}$ remains one of the most poorly constrained parameters of
the Cabibbo-Kobayashi-Maskawa (CKM) matrix~\cite{bb:CKM}.  Its
magnitude, $|V_{ub}|$, plays a central role in testing the
consistency of the CKM matrix in $B$ and $K$ meson decay processes.
Inconsistency would signal existence of new classes of fundamental particles or forces.   A precise determination of $|V_{ub}|$ has been the subject of considerable theoretical and experimental effort for well over a decade, and remains one of the highest priorities of flavor physics.  

Measurements of exclusive charmless semileptonic decays  $B\to X_u\ell\nu$~\cite{bb:cleoprl77y1996,bb:lange_cleo_exclusive,bb:Athar:2003yg,bb:BABARy2005,bb:BABARy2006,bb:BABARy2007,bb:BELLEy2007}  provide a route to the determination of  $|V_{ub}|$ with experimental and theoretical uncertainties complementary to the current inclusive techniques~\cite{bb:PDG_vub_minireview}. 
The primary challenge facing all measurements  of semileptonic $b\to u\ell\nu$ decay is separation of the signal from the much larger $b \to c\ell\nu$ background.  Exclusive $b \to u\ell\nu$ decays provide unique kinematic constraints for this purpose, particularly if a reliable estimate of the neutrino four-momentum is obtained. 

In addition to the experimental challenges facing the exclusive measurements, extraction of $|V_{ub}|$ requires 
knowledge of the form factors that govern the dynamics of these decays.  The form factors contribute both to the variation of rate, or shape, over phase space  and to the overall rate via the form factor normalization.  The form factors are inherently nonperturbative QCD quantities and considerable effort has been devoted to their calculation.
A major recent theoretical advance has come in the form of predictions from unquenched lattice QCD (LQCD) calculations 
~\cite{HPQCD_2006}, which provide predictions with uncertainties claimed to be at the 10\% level  for $B\to\pi\ell\nu$.    Light Cone Sum Rules (LCSR) techniques also provide predictions for the $B\to\pi\ell\nu$ ~\cite{ball_2005} and $B\to\rho\ell\nu$ form factors~\cite{ball04}. 
 
For the extraction of $|V_{ub}|$, we consider semileptonic decays to the pseudoscalar final states $\pi^-\ell^+\nu$ and $\pi^0\ell^+\nu$, for which we have the unquenched lattice QCD predictions.
We also investigate the decays to vector final states $\rho^-\ell^+\nu$, $\rho^0\ell^+\nu$, and $\omega\ell^+\nu$.  An accurate determination of the decay rate to vector final states is necessary for quantifying cross feed into the $B\to\pi\ell\nu$ rate that is ultimately used to extract $|V_{ub}|$. 

In addition, we search for $B^+\to\eta\ell^+\nu$ and 
$B^+\to\eta^{\prime}\ell^+\nu$. Because semileptonic decays involve only a single hadronic current, 
they additionally offer a probe of the QCD phenomena underlying fully 
hadronic decay.  In particular, the $B\to\eta\ell^+\nu$ and $B\to\eta^{\prime}\ell^+\nu$ decays 
probe the singlet or QCD anomaly component of the $\eta^{\prime}$ meson~\cite{etap:FKS}.  
This component may be responsible for the unexpectedly high branching 
fractions observed in $B \to \eta^{\prime} X_s$ decays 
\cite{etap:cleo:etapxs,etap:babar:etapxs,etap:belle:etapxs}.  
 A  measurement of $\mathcal{B}(B^+\to\eta\ell^+\nu)$ was included in
previous CLEO studies~\cite{bb:Athar:2003yg},
 and in 2006, \textsc{BaBar} released preliminary upper limits for both 
$B\to\eta\ell\nu$ and $B\to\eta^{\prime}\ell\nu$ branching fractions~\cite{BaBar_eta_ichep06}.

This paper describes the CLEO studies of these seven exclusive $b\to u\ell\nu$ decay modes. The excellent hermeticity of the CLEO detector and the symmetric $e^+e^-$ beam configuration of the Cornell Electron Storage Ring (CESR) allows for precise determination of the neutrino four-momentum. In contrast to asymmetric $B$ factories, the lack of a center of mass boost along the beam direction enhances the fraction of $\Upsilon(4S)\to B\bar{B}$ events in which all particles are contained in the detector fiducial region.  This allows more stringent selection criteria to be placed on the reconstructed neutrino that improve neutrino resolution without degrading efficiency.  This ultimately results in a cleaner separation between signal and background that enables measurements of exclusive semileptonic rates at CLEO to be competitive with higher statistics analyses carried out at asymmetric $B$ factories.

We study all seven decay modes simultaneously, extracting independent measurements in subregions of phase space in order to minimize {\em a priori} dependence on form factor shape predictions.  While the general method is similar to the previous CLEO study~\cite{bb:Athar:2003yg}, the current analysis further reduces the sensitivity to the decay form factors by decreasing the lower bound on the charged-lepton momentum requirement for the vector modes from $1.5$ GeV/$c$ to $1.0$ (1.2) GeV/$c$ for electrons (muons) and, in addition, measuring partial rates for vector modes in bins of lepton decay angle.  Furthermore, our measurements of the rates over the $\rho\ell\nu$ phase space allow us to test the validity of form factor shape calculations in that mode. 

To increase signal efficiency, we have also modified the phase-space binning in the region of high meson recoil momentum, where continuum $e^+e^-\to q\bar{q}$ backgrounds contribute.  

This analysis utilizes a total of $15.4 \times 10^6$ $B\overline{B}$ decays obtained at the $\Upsilon(4S)$, a relative increase of 60\% from the previous CLEO analysis~\cite{bb:Athar:2003yg}. The results presented here supersede those of the previous analysis.

The paper is organized as follows.  Sections~\ref{sec:exclusive_intro} and~\ref{sec:anomoly_intro}
discuss the role of form factors for extraction of $|V_{ub}|$ and the relation of the QCD anomaly to the
$B\to\eta\ell\nu$ and $B\to\eta^\prime\ell\nu$ branching fractions.  Section~\ref{sec:recon} describes our reconstruction technique.
Section~\ref{sec:extraction_fit_intro} describes our fitting method,
fit components, and fit results. Section 
\ref{sec:systematics}  describes the systematic
uncertainties. Section~\ref{sec:discussion} interprets the results. 
Throughout the paper, charge conjugate modes are implied.
A more detailed description of the analysis technique can be found in Reference~\cite{bb:mrs43_thesis}. 

\section{Exclusive Charmless Semileptonic Decays}
\label{sec:exclusive_intro}

\begin{figure}[b]
\begin{center}
\includegraphics[width=3in]{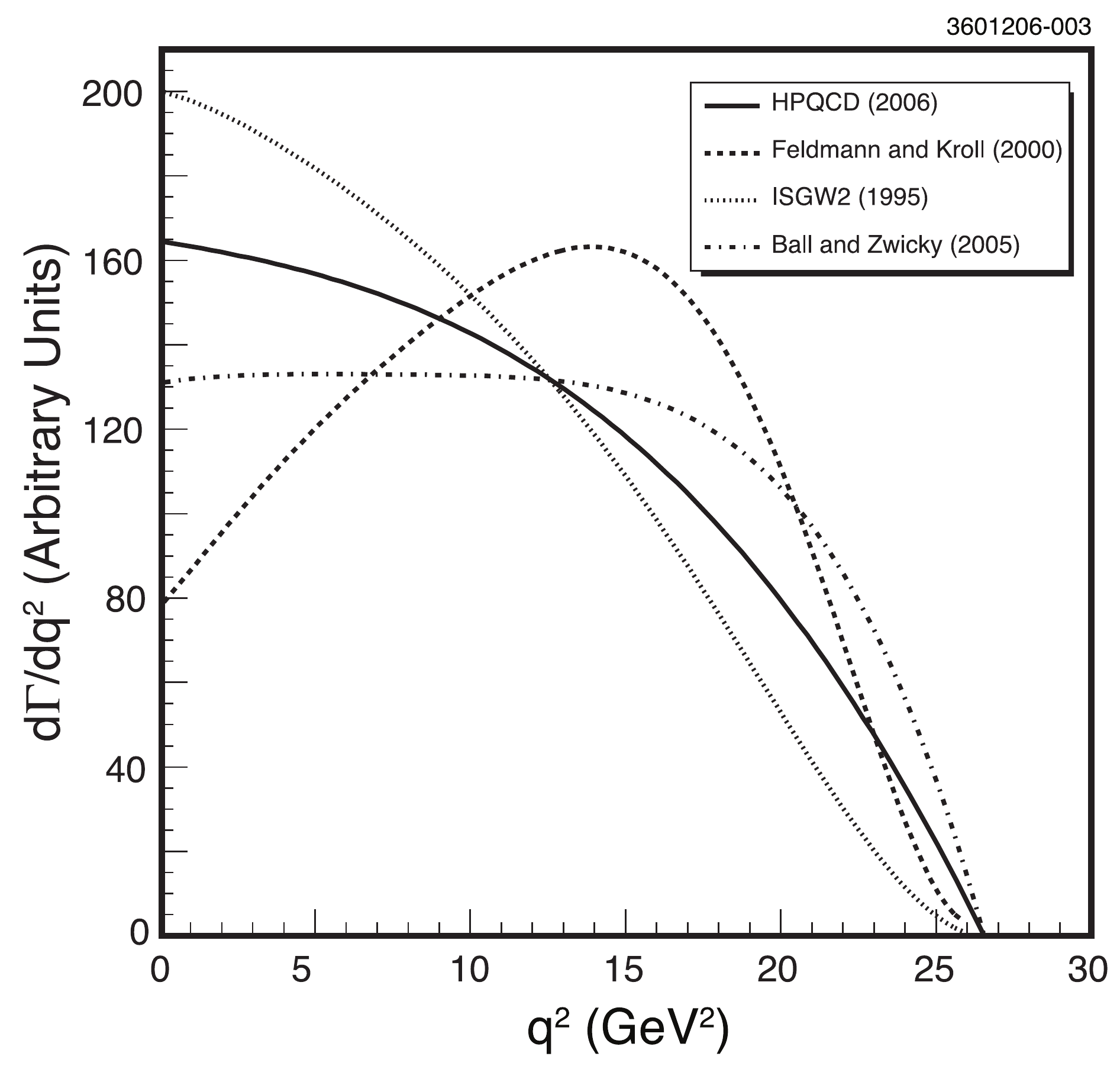}
\caption{\label{fig:pi_formfactors}Predictions for $d\Gamma(B\to\pi\ell\nu)/dq^2$ for a variety form factor calculation techniques~\cite{HPQCD_2006,spd,isgw2,ball_2005} illustrate
the range of variation of the predicted $q^2$--dependence. }
\end{center}
\end{figure}

\begin{figure*}[tb]
\begin{center}
\includegraphics[width=6in]{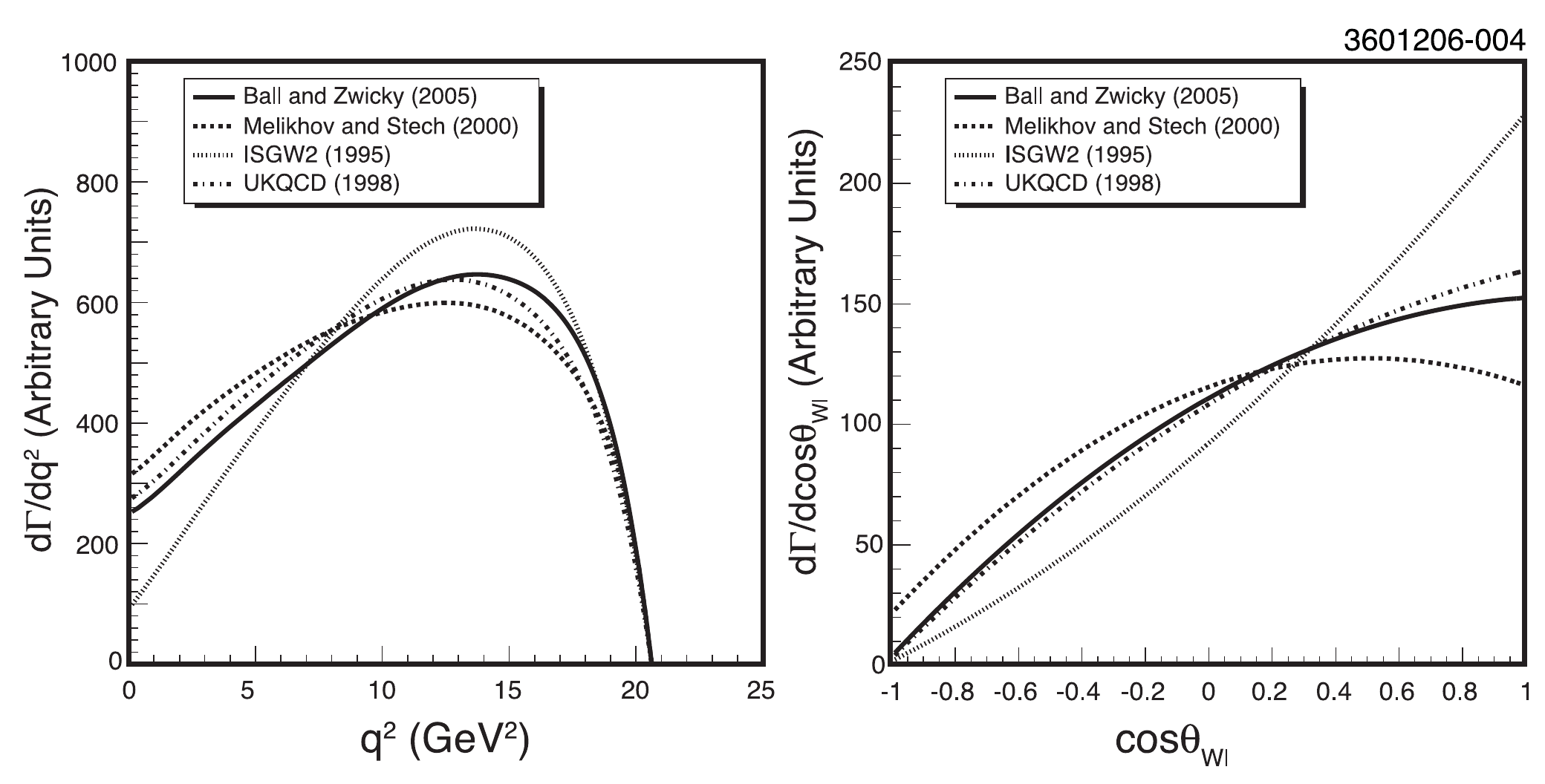}
\caption{\label{fig:rho_formfactors}Predictions for $d\Gamma(B\to\rho\ell\nu)/dq^2$ (left) and $d\Gamma(B\to\rho\ell\nu)/d\cos\theta_{Wl}$ (right)  for a variety of  form factor calculation techniques~\cite{ball04,meli,isgw2,ukqcd98} illustrate the range of variation of the predicted $q^2$ and $\cos\theta_{Wl}$ dependence.}
\end{center}
\end{figure*}

For $B\to V_u\ell\nu$, where $V_u$ is a charmless vector meson, the partial width is
\begin{equation}
\frac{d\Gamma}{dq^2\,dC_\theta} =\kappa kq^2 \left[ S^2_\theta H_0^2 +
\frac{C_-^2H_+^2 +  
C_+^2H_-^2}{2} \right].
\end{equation}
Here, $\kappa= \frac{|V_{ub}|^2 G_F^2}{128\pi^3M_B^2}$,  $k$ is the $V_u$ momentum, $q^2$ is the mass squared of the virtual $W$ ($W^*$), $C_\theta$ ($S_\theta$) is the cosine (sine) of the angle $\theta_{Wl}$ between the charged lepton in the $W^*$ rest frame and the $W^*$ in the $B$ rest frame, and $C_\pm=1\pm C_\theta$.  
$H_\pm$ and $H_0$ are the magnitudes of the $W$~helicity amplitudes, which can be expressed in the massless lepton limit in terms of three $q^2$--dependent form factors~\cite{Gilman_and_Singleton}, $g$, $a_+$, and $f$, as
\begin{eqnarray}
&&\left|H_\pm\right|^2 = \left[f\left(q^2\right)\mp 2 M_B k g\left(q^2\right)\right] ^2
\label{eq:vechpm} \\
~\nonumber\\
&&\left|H_0\right|^2 = \frac{M_{B}^4}{4q^2M_{V_u}^2}\times\nonumber\\
&&\Bigg[\left(1-\frac{M_{V_u}^2+q^2}{M_B^2}\right)f(q^2)+
4k^2a_+(q^2)\Bigg]. \label{eq:vech0}
\end{eqnarray}

For a final state pseudoscalar meson $P_u$, $H_{\pm}=0$, and the rate depends on a single form factor
$f_+(q^2)$:
\begin{equation}
\frac{d\Gamma\left(B\to P_u\ell\nu\right)}{dq^2}=\left|V_{ub}\right|^2\frac{G_F^2}{24\pi^3}k^3\left|f_+\left(q^2\right)\right|^2.
\label{eq:dgammadq2}
\end{equation}

The structure of the differential decay rates allows us to
draw some general conclusions regarding the properties of the semileptonic decay
modes studied here.   For the $\rho(\omega)\ell\nu$ transitions,
the left-handed, $V-A$, nature of the charged current at the quark level
manifests itself at the hadronic level as $|H_-|>|H_+|$.
 The $H_-$ contribution is also expected to dominate the $H_0$ contribution,
leading to a forward-peaked distribution for $\cos\theta_{Wl}$.  The pseudoscalar
modes exhibit a $\sin^2\theta_{Wl}$ dependence, independent of the
form factor since the rate is independent of $\theta_{Wl}$.  They also depend on an extra
factor of $p_{X_u}^2$, which suppresses the rate near $q^2_{\text{max}}$ ($p_{X_u}=0$).
Taken together, these two effects give the pseudoscalar modes a softer
charged lepton momentum spectrum than the vector modes.

Recent results for $f_{+}(q^2)$ from lattice calculations with dynamical quarks~\cite{HPQCD_2006} provide a marked theoretical advance in form factor calculations.  However, significant theoretical uncertainties still remain in the $q^2$ dependence of the form factors, particularly in the non-$\pi\ell\nu$ decay modes.  Figures~\ref{fig:pi_formfactors}~and~\ref{fig:rho_formfactors} illustrate the variation of form factor predictions for a variety of theoretical techniques.  The effect of these uncertainties on the total rate measurements can be mitigated by measuring partial rates in several regions of phase space.

In the $\pi\ell\nu$ modes the rates 
extracted in our chosen $q^2$ intervals will be largely independent of the assumed
form factor shapes.  In the
vector modes, however, the three form factors interfere and uncertainties in this
interference, particularly for $\cos\theta_{Wl}<0$,
lead to a form factor systematic uncertainty.  To reduce this uncertainty, we have included the 
$\cos\theta_{Wl}<0$ data in the fit, rather than relying on our assumed form factors to extrapolate
into this region.   There is substantial cross feed from the $\cos\theta_{Wl} < 0$ region of $B\to\rho\ell\nu$ phase space into the reconstructed $B\to\pi\ell\nu$ decays.  As a result,
the  the largest form factor systematic uncertainty in the previous  CLEO  $B\to\pi\ell\nu$  measurement arose from the $B\to\rho\ell\nu$ mode.  Our improved measurements over a larger region of phase space  better constrain the $B\to\rho\ell\nu$ rate and reduce this systematic uncertainty by roughly an order of magnitude.

\section{QCD Anomaly in $B\to\eta^{(\prime)}\ell\nu$}
\label{sec:anomoly_intro}

As mentioned in the introduction,
we can check for consistency and probe for new information in quark 
symmetries by measuring the $B\to \eta\ell\nu$ and  $B\to \eta^\prime\ell\nu$ branching fractions. 
Branching fraction measurements for $B \to \eta^{\prime} X_s$~\cite{cleo_etapk_1997,etap:cleo:etapxs,etap:babar:etapxs,etap:belle:etapxs}
  yield larger than expected results, indicating extra couplings in the $B\to\eta^{\prime}$ form factor. 
Measurements at CLEO of  $\eta^\prime$ production in $\Upsilon(1S)$ decays~\cite{Artuso:2002px} limit the contribution of the $\eta^\prime\to g^*g$ form factor to the anomalously high $B\to\eta^\prime X_s$ rate.  This unexpected rate may be accounted for by the additional
 QCD Anomaly~\cite{QCDAnomalyDefinition} contribution to the $\eta^{\prime}$ axial vector current, which
can be cleanly measured in semileptonic decays~\cite{etap:CSK}.
The QCD Anomaly provides additional
gluon couplings to singlet states, but no additional couplings to non-singlet states.

The $SU(3)$ octet of the {\it u}, {\it d}, and {\it s} quarks contain 
the  $\pi^-$, $\pi^+$, $K^0$, $K^-$, $K^+$ mesons as well as sums of
the $\pi^0$ with the octet triplet 
$\eta^8\equiv \frac{1}{\sqrt{6}}(\bar{u}u+\bar{d}d-2\bar{s}s)$. The singlet 
of this symmetry is 
$\eta^0 \equiv \frac{1}{\sqrt{3}}(\bar{u}u+\bar{d}d+\bar{s}s)$.
The physical $\eta$ and $\eta^{\prime}$ mesons appear to be primarily
 made up of
 combinations of the $\eta^0$ singlet state and the $\eta^8$ triplet state.
Since the $\eta$ mass is so much closer to the $K$ masses, it is 
primarily $\eta^8$. The $\eta^{\prime}$ is mostly the $\eta^0$ singlet, which 
possesses the strong gluon couplings in the axial vector current. 
A measurement of the ratio of $\mathcal{B}(B\to\eta^\prime\ell\nu)$ to $\mathcal{B}(B\to\eta\ell\nu)$ provides experimental input needed to extract the strength of these gluon couplings.

Once the sizes of these couplings have been determined for a particular model
using the semileptonic $B\to\eta\ell\nu$ and $B\to\eta^{\prime}\ell\nu$ 
decays, the same form factor parameterization can
be used to check for consistency with the $B\to\eta^{\prime}X_s$ decays (see section \ref{sec:eta_discussion}.) 

\section{Event Reconstruction and Selection}
\label{sec:recon} 

This study utilizes the full 15.5~fb$^{-1}$ set of data collected at the 
$\Upsilon(4S)$  with the 
CLEO~II~\cite{Kubota:1991ww}, CLEO~II.5~\cite{Hill:1998ea} and CLEO~III~\cite{Peterson:2002sk} detectors at CESR. 
The analysis rests upon associating  the missing energy and 
momentum in each event with the neutrino four-momentum, an
approach enabled by the excellent hermeticity and resolution of the CLEO detectors.  
Charged particles  are  detected over at least 93\% of the total solid angle for all three detector configurations, and are measured in a solenoidal magnetic field with a momentum resolution of 0.6\% at 2 GeV/$c$.  Photons and electrons are detected in a CsI(Tl) electromagnetic calorimeter that covers 98\% of the $4\pi$ solid angle.  A typical $\pi^0$ mass resolution is 6 MeV$/c^2$.  Unless otherwise noted, all kinematic quantities are measured in the laboratory frame.

Electrons satisfying $p>200\text{ MeV}/c$  are identified over 90\% of the solid angle by using the
ratio of energy deposited in the calorimeter to track momentum in conjunction with specific ionization ($dE/dx$) information from the main drift chamber, shower shape information, and the difference between extrapolated track position and the cluster centroid.  Depending on detector configuration, either time-of-flight (CLEO~II/II.5) or 
Ring Imaging \v Cerenkov~\cite{Artuso:2005dc} (CLEO~III) measurements provide additional $e^\pm/K^\pm$ separation.

Particles  in the polar angle range $|\cos\theta|<0.85~(0.65)$  that register hits in counters beyond five
interaction lengths are
accepted as signal muons with the CLEO~II/II.5 (CLEO~III) detector configurations. Those with $|\cos\theta|<0.71$ and hits between three and five interaction lengths  are used in a multiple-lepton veto, discussed below.   

We restrict signal electron  and muon candidates  to the momentum interval $1.0 < p < 2.8 \text{ GeV}/c$ and $1.2 < p < 2.8 \text{ GeV}/c$, respectively.    The lower limits are determined by excessive background contributions for electrons, and by identification efficiency for muons.
We also require that the lepton candidate tracks have signals from at  least 40\% of their potential drift chamber layers and are consistent with originating from the beam spot.

Momentum-dependent lepton identification efficiencies and the rates for hadrons being misidentified as a lepton (``fake rates'') for both electrons and muons are measured in data by examining the rate at which the decay products from cleanly identified $\Lambda\to p \pi^-$, $K^0_S\to\pi^+\pi^-$, and $D^{*+}\to\pi^+D^0\to\pi^+K^-\pi^+$ decays satisfy lepton identification criteria.  Within the signal electron fiducial and momentum regions, the identification efficiency exceeds 90\% while the fake rate is about $0.1\%$.  For signal muons above 1.5 GeV$/c$, the identification efficiency also averages above 90\%. For muons below 1.5 GeV$/c$,  the five interaction length requirement causes the efficiency to fall rapidly to about 30\% at the lowest momentum of 1.2 GeV$/c$.  The muon fake rate is about 1\%.  The measured efficiencies and fake rates are employed in all Monte Carlo (MC) simulations.

Each charged track that is not identified as an electron or muon is assigned the most probable of
the pion, kaon or proton mass hypotheses.  The probability for each mass hypothesis is formed
from an identification likelihood, based upon the specific ionization measurements
in the drift chamber and either the time-of-flight or \v Cerenkov photon angle  
measurements, and the relative production fractions for pions, kaons and protons at that momentum in generic $B$ meson decay.  

To reconstruct the undetected neutrino we associate the neutrino four-momentum
$p_\nu$ with the missing four-momentum $p_{\mathrm{miss}}$.
In the process $e^+ e^-
\to \Upsilon(4S) \to B\bar{B}$, the total energy of the beams is imparted to
the $B\bar{B}$ system; at CESR that system is very nearly at rest because beam energies
are symmetric and the beam crossing angle is small ($\approx$~2~mrad). 
The missing four-momentum in an event is given by $p_{\mathrm{miss}} =
(E_{\mathrm{miss}},\vec{p}_{\mathrm{miss}})= 
 p_{\mathrm{total}} - \sum p_{\mathrm{charged}} - \sum
 p_{\mathrm{neutral}}$, where the event
four-momentum $p_{\mathrm{total}}$  is known from the energy
and crossing angle of the CESR beams and $p_{\mathrm{charged}}$ and $p_{\mathrm{neutral}}$ are the four-momenta of charged and neutral particles explicitly reconstructed by the detector.  Charged
and neutral particles pass selection criteria designed
to achieve the best possible $|\vec{p}_{\mathrm{miss}}|$ resolution by
balancing the efficiency for detecting true particles against the
rejection of false ones. 

For the charged four-momentum sum $\sum
p_{\mathrm{charged}}$, optimal selection is achieved with topological criteria that minimize
multiple-counting resulting from low-momentum tracks that curl in the
magnetic field, charged particles that decay in flight or interact within
the detector, and spurious tracks.  Tracks that are actually segments of a single low transverse
momentum ``curling'' particle can be mis-reconstructed as two (or more) separate outgoing tracks with opposite charge and roughly equal momentum.  These tracks are identified by selecting oppositely charged
track pairs whose innermost and outermost diametric radii each match
within 14 cm and whose separation in $\phi$ is within
180$^\circ \pm 20^{\circ}$.  We choose the track segment that
will best represent the original charged particle based on track
quality and distance of closest approach information.
We employ similar algorithms to identify particles that
curl more than once, creating three or more track segments.
We also identify tracks that have scattered or decayed in the drift
chamber, causing the original track to end and one or more tracks to
begin in a new direction.  We keep only the track segment with the majority of its hits
before the interaction point.  Spurious tracks are identified by their low hit
density and/or low number of overall hits, and rejected.

For the neutral four-momentum  sum, $\sum  p_{\mathrm{neutral}}$, clusters
resulting from the interactions of charged hadrons must be avoided.  As a
first step, calorimeter showers passing the standard CLEO
proximity-matching (within 15 cm of a charged track) are eliminated.
Optimization studies also revealed that all showers under 50 MeV should be
eliminated.  The processes that result in separately reconstructed
showers (``splitoffs'') within about~25$^\circ$ of a proximity-matched
shower  tend to result in an energy distribution over the $3\times3$
central array of the splitoff shower that ``points back'' to the core
hadronic shower.  We combine this pointing information with the ratio of
energies in the $3\times3$ to $5\times5$ arrays of crystals, whether the
shower forms a good $\pi^0$, and the MC predictions for relative
energy spectra for true photons versus splitoff showers to provide an optimal
suppression of the splitoff contribution.

Signal MC events show a $|\vec{p}_{\text{miss}}|$ resolution of $\approx 0.1$ GeV/$c$ and  
an  $E_{\text{miss}}$ resolution of $\approx 0.2$ GeV after all analysis cuts.

To further enhance the association of the missing momentum with an
undetected neutrino in our final event sample, we require that
the missing mass, $M^2_{\text{miss}} \equiv E^2_{\text{miss}} - |\vec{p}_{\text{miss}}|^2$,
be consistent within resolution with a massless neutrino. Specifically, since the
$M^2_{\text{miss}}$ resolution scales as $\sim 2E_{\text{miss}}/\sigma_{E_{\text{miss}}}$, 
for each exclusive $B\to X_u\ell\nu$ mode we reconstruct, we
 require $M^2_{\text{miss}}/2E_{\text{miss}}$, with typical ranges of
  $-0.5 < M^2_{\text{miss}} / 2E_{\text{miss}} < 0.3\ \text{GeV}$. The criterion was
 optimized for each decay mode using independent samples of MC simluations for signal and background processes.
 
Association of the missing four-momentum with the neutrino
four-momentum is only valid if the event contains no more than one
neutrino and all true particles are detected.  For events with
additional missing particles or double-counted particles, the signal
modes tend not to reconstruct properly, while background processes tend
to smear into our sensitive regions.  Hence it is worthwhile to reject
events in which missing particles are likely.   
We exclude events containing more than one identified lepton, which indicates an
increased likelihood for multiple neutrinos.  

We require the angle of the missing momentum with respect to the beam axis satisfy $|\cos\theta| < 0.96$.  In addition to suppressing events in which particles exited the detector along the beam line, this requirement also suppresses events resulting from a collision of radiated photons from the beams.  These two-photon events are background to $B\bar{B}$ events and are typically characterized by a large missing momentum along the beam axis. 

A nonzero net charge $Q$
indicates at least one missed or doubly counted charged particle.  The 
$Q=0$ sample offers the greatest purity.
For the pseudoscalar modes, use of the $|Q|=1$ in addition to the
$Q=0$ sample offers a statistical
advantage.  Events from the subset in which a very soft pion from a $D^*$ decay 
was not reconstructed in particular have only a modestly distorted missing
momentum.  While we retain the  $|Q|=1$ sample,
we treat it separately from $Q=0$ sample in reconstruction and fitting
so that the statistical power of the latter sample, with its  better signal to
background ratio, is not diluted.  For the vector modes, which have the poorest signal to background ratios, we require $Q=0$ because systematic errors associated with the normalization of the larger background in the $|Q|=1$ sample outweigh the statistical benefits.

After our selection criteria, the remaining $b\to c\ell\nu$ background
events are dominated by events that contain
either a $K_L$ meson or an additional neutrino
that is roughly collinear with the signal neutrino.

With an estimate of the neutrino four-momentum in hand, we can employ full
reconstruction of our signal modes.
Because the resolution on $E_{\text{miss}}$ is so much larger than that for 
$|\vec{p}_{\text{miss}}|$, we use
$(E_\nu,\vec{p}_\nu)=(|\vec{p}_{\text{miss}}|,\vec{p}_{\text{miss}})$ for
full reconstruction.  
The neutrino combined with the signal charged lepton ($\ell$) and hadron ($h$) should
satisfy, within resolution, the constraints on energy, $\Delta E \equiv
(E_\nu+E_\ell+E_h)-E_{\text{beam}} \approx 0$, and on momentum, $M_{h\ell\nu}
\equiv [E_{\text{beam}}^2 - |\alpha\vec{p}_\nu+\vec{p}_\ell+\vec{p}_h|^2]^\frac12 \approx
M_B$, where $\alpha$, a momentum scaling factor, is chosen such that  $\alpha E_\nu+E_\ell+E_h - E_{\text{beam}}=0$.  The neutrino
momentum resolution dominates the $\Delta E$ resolution, so the momentum
scaling corrects for the mismeasurement of the magnitude of the neutrino momentum
in the $M_{h\ell\nu}$ calculation.  This correction also reduces the correlation between  $M_{h\ell\nu}$ and $\Delta E$ mismeasurement.  Uncertainty in the neutrino direction remains as the
dominant source of smearing in this mass calculation.  

We reconstruct $q^2=M_{W^*}^2=(p_\nu+p_\ell)^2$ for each decay from the
reconstructed charged lepton four--momentum and the missing momentum.  In addition
to using the scaled reconstructed momentum $\alpha\vec{p}_\nu$ described
above, the direction of the missing momentum is changed through the
smallest angle consistent with forcing $M_{h\ell\nu}= M_B$.  This procedure
results in a $q^2$  resolution of 0.3 GeV${^2}$, independent of $q^2$. 

A signal charged $\pi^+$ candidate must have the pion hypothesis as its most probable particle ID outcome, and it must not be a daughter in any reconstructed $K_s$ decay. 
A $\pi^0$ candidate must have a $\gamma\gamma$ mass within 2 standard
deviations of the $\pi^0$ mass and an energy greater than $250\text{ MeV}$.
Each daughter shower must have an energy greater than $30\text{ MeV}$.  Combinatoric
background levels outstrip efficiency gains below either of these requirements.

Reconstructed $\rho^0\to \pi^+\pi^-$
and $\rho^+\to \pi^+\pi^0$ candidates are accepted within 285 MeV$/c^2$ of the nominal $\rho$ mass to accommodate the broad $\rho$ resonance width.   The  $\pi\pi$ mass ranges are divided
into three 95~MeV$/c^2$ intervals for the remainder of the reconstruction and fitting.
We accept $\omega$ candidates, reconstructed  via its $\pi^+\pi^-\pi^0$ decay, within 30 MeV$/c^2$ of the nominal $\omega$ mass, and, similar to the $\rho$ modes, the $\pi^+\pi^-\pi^0$ invariant mass region is divided into three equal intervals of 10~MeV$/c^2$.  This division in both modes makes the fit sensitive to signal through rough $\rho/\omega$ line shape information.  Except where noted, plots in this paper will include data from only the invariant mass interval centered on the resonance mass although data from all three regions are included in the fit.

We reconstruct $\eta$ in both the $\gamma\gamma$ and the $\pi^+\pi^-\pi^0$ decay modes.
For $\gamma\gamma$, we require the reconstructed mass to be within
 2 standard deviations (about 26 MeV$/c^2$) of the $\eta$ mass and each shower
to have $\cos\theta<0.81$ relative to the beam pipe.  Showers within a reconstructed $\pi^0$
are vetoed.
For the $\eta\to \pi^+\pi^-\pi^0$ submode,   the $\pi^0$ must be within 2 standard deviations of
the $\pi^0$ mass and statisfy $E_{\pi^0}>225$ MeV. To reduce combinatoric backgrounds, the 
invariant mass of the two charged pion tracks must satisfy  $M_{\pi\pi} < M_{\eta} - M_{\pi^0}+0.05$ GeV$/c^2$.

\begin{figure}[tb]
\begin{center}
\includegraphics[width=3.2in]{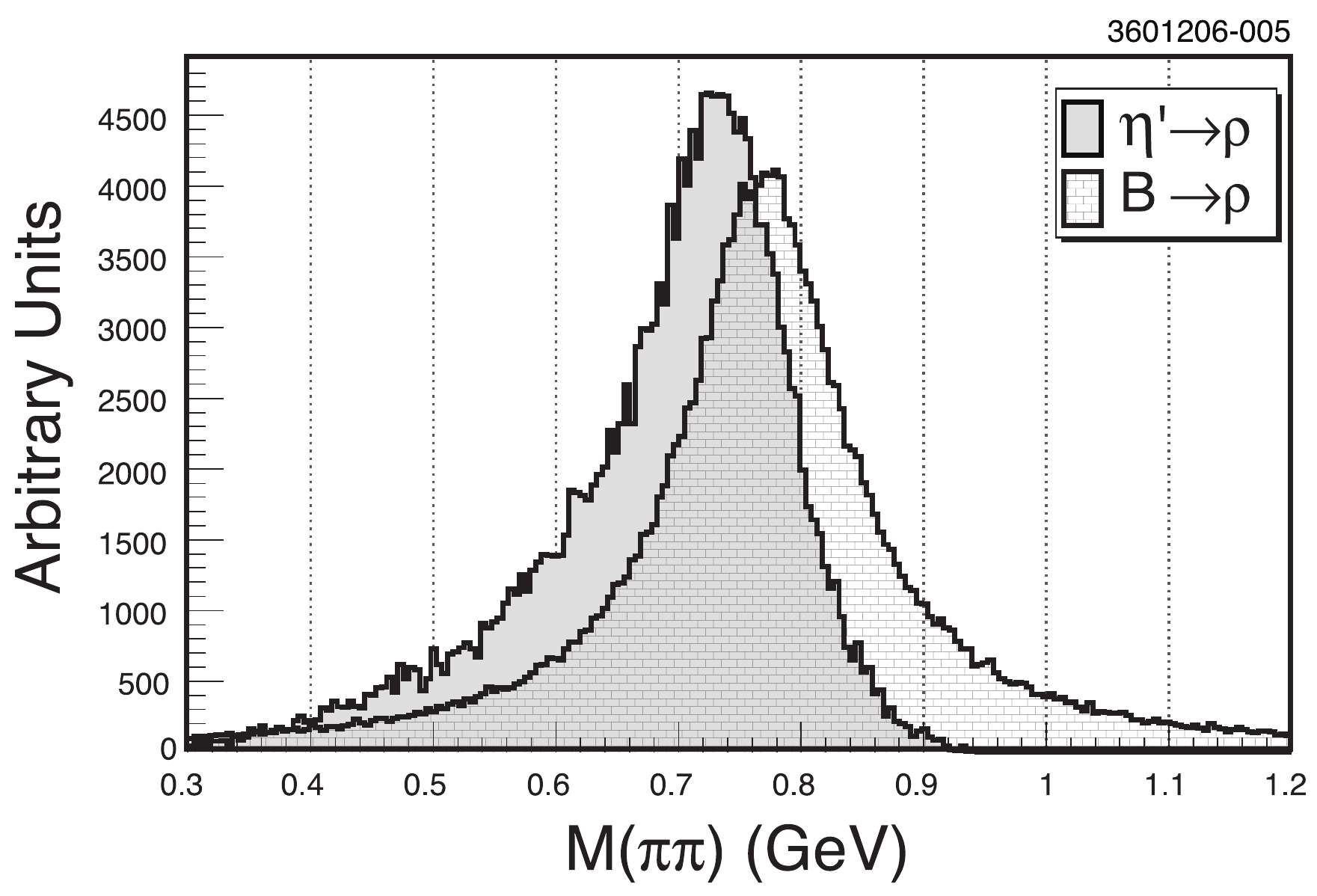}
\caption{\label{fig:rho_line_shape} The shape of the mass of the $\rho$ in $B\to\rho\ell\nu$ and
$\eta^{\prime}\to\rho\gamma$. Limited phase space in  
$\eta^{\prime}\to\rho\gamma$ produces the asymmetry about the nominal $\rho$ mass in this mode.
Histograms are normalized to equal area.}
\end{center} 
\end{figure}

The $\eta^{\prime}$ mode is reconstructed in $\eta^{\prime}\to\rho\gamma$ and $\eta^{\prime}\to\eta\pi\pi;~\eta\to\gamma\gamma$.  
 Selection criteria are optimized separately based on the event net 
charge and $q^2$ where we distinguish between $q^2>10~\text{GeV}^2$ and $q^2<10~\text{GeV}^2$.  
Here we specify the requirements for the $Q=0$ and $q^2<10~\text{GeV}^2$ sample, for which we have the greatest sensitivity.
For $\eta^{\prime}\to\rho\gamma$, combinatoric backgrounds are reduced by 
requiring 
$\theta_{\pi\gamma}$, the angle between a pion and the photon in the $\rho$ 
rest frame, to satisfy $|\cos\theta_{\pi\gamma}|<0.925$.  Photons 
from $\pi^0$ candidates were vetoed.  We accept $\rho$ candidates within the mass range 0.3 to 0.9 GeV$/c^2$, where the range is subdivided into 
four roughly equal mass bins.
Note that phase space restrictions in $\eta^{\prime}\to\rho\gamma$ 
distort the $\rho$ line shape at high $\pi\pi$ mass (Figure \ref{fig:rho_line_shape}). 
The line shape falls rapidly to zero at $M_{\pi\pi}=0.9$~GeV$/c^2$, so the largest mass
bin, $0.78$~GeV/$c^2~-~0.9$~GeV$/c^2$, is expected to have relatively little signal.  That mass region is primarily used in the fit to constrain the background level at lower mass.  The reconstructed $\eta^{\prime}$ mass must be within 2.5 standard deviations of the nominal $\eta^\prime$ mass.   

For the $\eta^{\prime}\to\eta\pi\pi~(\eta\to\gamma\gamma)$ mode,
the $\eta$ daughter photons must satisfy the same criteria as the $\eta$ in 
$\eta\ell\nu$.  We also require $E_\eta> 300$ MeV.  
Finally, we require the reconstructed  $\eta^{\prime}$ and $\eta$ masses 
to satisfy $(\sigma^2_{\eta^{\prime}}+\sigma^2_{\eta})^{1/2}<3.75$, where 
$\sigma_{\eta(\eta')}$  is the number of standard deviations  of the reconstructed
mass away from the nominal mass.

We minimize multiple candidates per event to simplify statistical 
interpretation.  Within each $\pi$,  $\eta$ or $\eta^\prime\to\eta\pi^+\pi^-$ 
mode, we allow only one candidate within the region  $5.175\le M_{h\ell\nu}<5.2875$ GeV/$c^2$ and
$-0.75 <  \Delta E <0.25$ GeV used in the fit (Section~\ref{sec:extraction_fit_intro}).
Within a mode, multiple candidates are resolved by choosing the one with the smallest $|\Delta E|$. For the vector and the
 $\eta^\prime\to\pi^+\pi^-\gamma$ modes, the $|\Delta E|$ criterion combined
 with large combinatorics outside the signal region can induce a severe 
efficiency loss.  To mitigate that loss, we select the best candidate 
separately within each of the individual  $M_{\pi\pi}$ or $M_{3\pi}$ ranges. 
 The $|\Delta E|$ criterion induces only very slight peaking in backgrounds, which is modeled by the MC generated distributions used to fit the data.

\begin{figure}[tb]
\begin{center}
\includegraphics[width=2.7in]{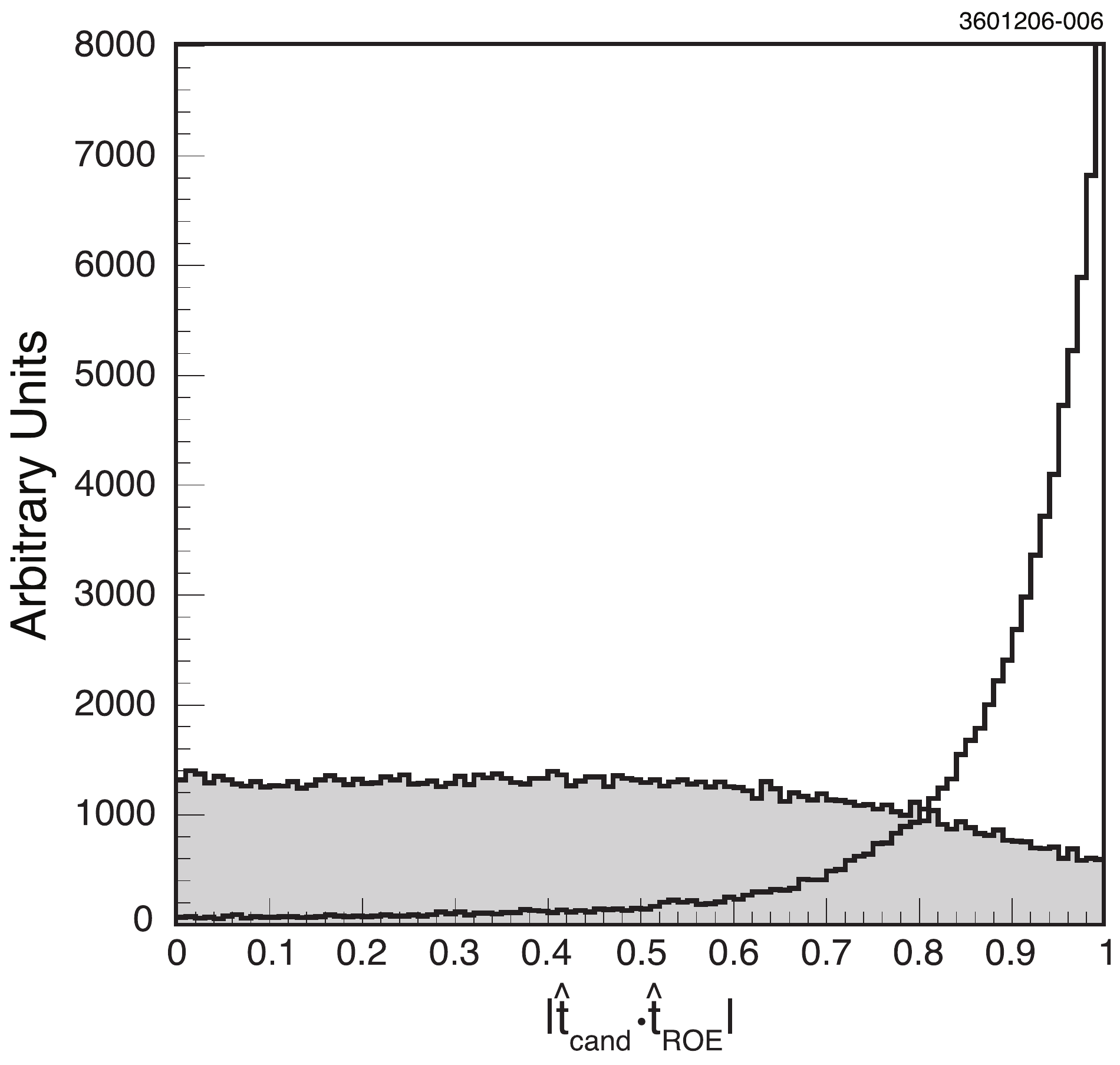}
\caption{\label{fig:thrust} A comparison
of $\left | \hat{t}_{\text{cand}} \cdot \hat{t}_{\text{ROE}} \right|$ for $B\to\pi\ell\nu$ signal MC (shaded) and off-resonance continuum data (open). }
\end{center} 
\end{figure}

\begin{figure}[tb]
\begin{center}
\includegraphics[width=2.7in]{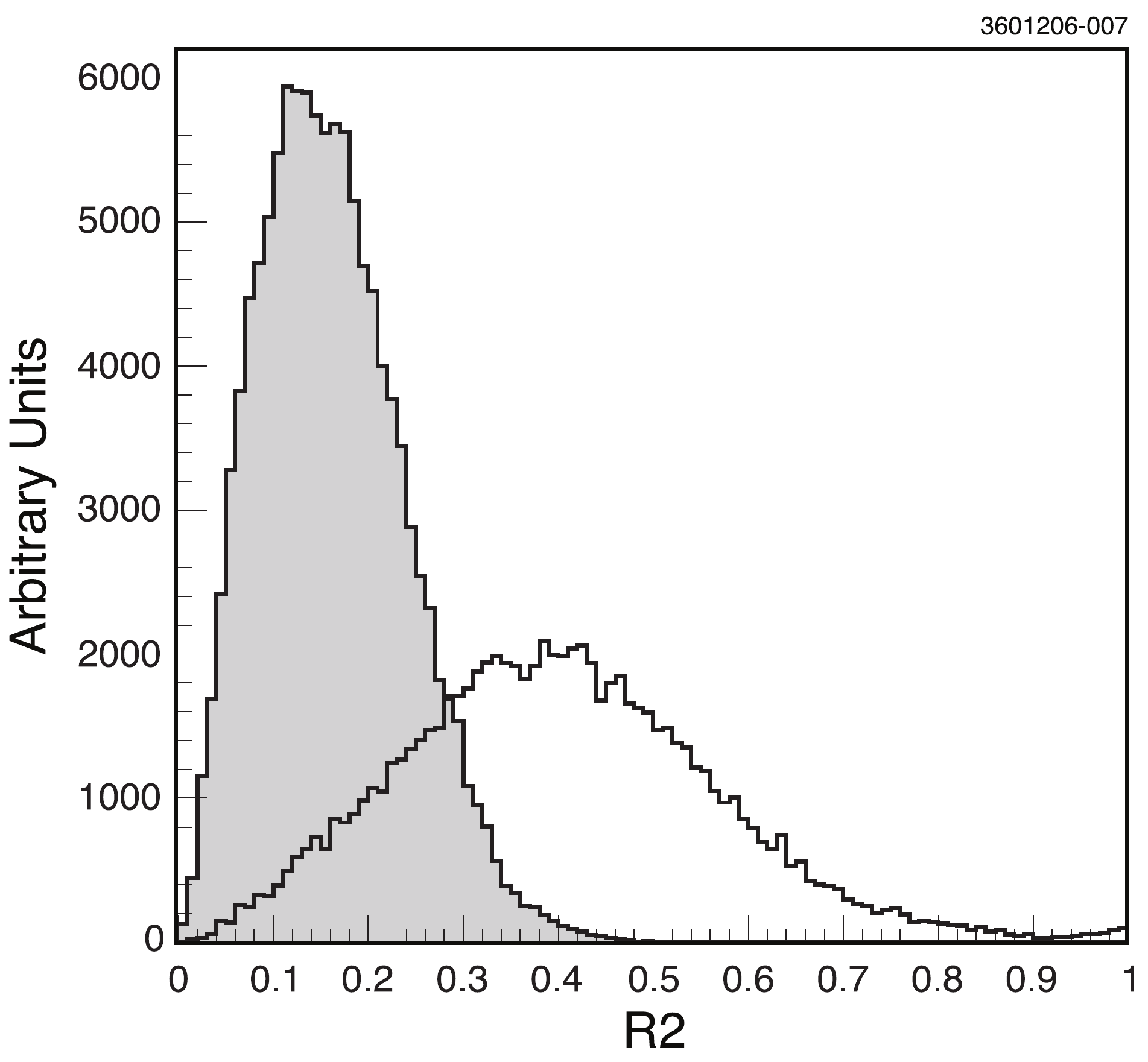}
\caption{\label{fig:r2}A comparison
of $R2$ for $B\to\pi\ell\nu$ signal MC (shaded) and off-resonance continuum data (open). }
\end{center}
\end{figure}

Backgrounds arise from the $e^+e^-\to q\bar{q}$ and
$e^+e^-\to \tau^+\tau^-$ continuum, fake
leptons, $b\to c\ell\nu$, and $B\to X_u\ell\nu$ modes other than the signal
modes.
Backgrounds from continuum processes make up approximately 75\% of the total
cross section at the $\Upsilon(4S)$ energy. The continuum processes typically produce two
collinear jets of hadrons, and the jet-like shape of the continuum events can be used to separate those events
from the isotropic $B\bar{B}$ resonance events. To do this, we  employ  
a Fisher Discriminant~\cite{fishorig} constructed from twelve variables that describe event shape.

The first input variable is $\left | \hat{t}_{\text{cand}} \cdot \hat{t}_{\text{ROE}} \right|$, where $\hat{t}_{\text{cand}}$ is the 
thrust axis obtained using  our candidate particles (excluding the neutrino), and
$\hat{t}_{\text{ROE}}$
that for the remaining particles in the event.  Continuum events typically have hadrons collimated into jets and therefore $\left | \hat{t}_{\text{cand}} \cdot \hat{t}_{\text{ROE}} \right|$ will peak near one (shown in Figure~\ref{fig:thrust}).  For $\Upsilon(4S)\to B\bar{B}$, the $B$ mesons are nearly at rest and the thrust directions from the two $B$ decays will be uncorrelated; therefore $\left | \hat{t}_{\text{cand}} \cdot \hat{t}_{\text{ROE}} \right|$ will be uniformly distributed.

We also use the ratio of the second to zeroth Fox-Wolfram moments~\cite{bb:fox_wolfram}, $R2\equiv H_2/H_0$.   The jet structure present in continuum events enhances the second moment and therefore the ratio $R2$ tends to one for jet-like continuum events and to zero for isotropic $B\bar{B}$ event (Figure~\ref{fig:r2}).

These two variables  provide the main discrimination between $B\bar{B}$ and continuum events. To gain further discriminating power we also consider the angle $\theta$ between the thrust axis of the entire event, $\hat{t}_{\text{event}}$, and the beam axis.  For $B\bar{B}$ events this variable will be randomly distributed, while for continuum events the $\hat{t}_{\text{event}}$ will align with the jet axis and be distributed with approximately the $1 + \cos^2\theta$ dependence of the cross section for $e^+e^-\to q\bar{q}$.

The remaining nine variables track the momentum flow of the event into nine forward-backward cones about the event thrust axis,  each spanning 10$^\circ$ in polar angle from the thrust axis. We optimize the continuum suppression criterion for each mode (and each subregion of phase space for each mode) separately.  Due to their jet-like nature, continuum backgrounds typically reconstruct with low $q^2$ and can be largely isolated at $q^2 < 2$ GeV$^2$.  The continuum suppression has an efficiency of roughly 15\% for continuum background and 90\% for signal $B\to\pi\ell\nu$ events. 

The signal selection efficiencies, averaged over phase space, for the restricted signal region range from $\approx 0.5-4\%$ and are summarized in Table~\ref{tab:eff}.  For cases such as $\eta$ and $\eta^\prime$, where we only reconstruct a fraction of the decay modes, the branching fractions~\cite{pdg06} to the modes we reconstruct are included in the efficiency.

\begin{table}
\caption{The reconstruction efficiency and fit signal yield for each final state. Both numbers are integrated over all phase space, using the nominal form factors for our fit (Section~\protect\ref{sec:extraction_fit_intro}).  The efficiencies include the branching fractions for the reconstructed modes of the signal hadron.  For the case of $\rho\ell\nu$ and $\omega\ell\nu$ the efficiency and yield in the central mass bin are reported.  Errors on the yield are statistical only.}
\label{tab:eff}
\begin{tabular}{@{\hspace{5pt}}c@{\hspace{5pt}}@{\hspace{5pt}}c@{\hspace{5pt}}c@{\hspace{5pt}}}\hline\hline
{\bf Decay Mode} & {\bf Efficiency} & {\bf Fit Yield [Events]}\\ \hline 
$B\to\pi^\pm\ell\nu$ & 4.3\% & $179\pm20$  \\
$B\to\pi^0\ell\nu$ & 2.6\% & $60\pm 6$ \\
$B\to\eta\ell\nu$ & 1.0\% & $14\pm 7$ \\
$B\to\rho^\pm\ell\nu$ & 0.8\% & $71\pm 9$ \\
$B\to\rho^0\ell\nu$ & 1.6\% & $79\pm 10$ \\
$B\to\omega\ell\nu$ & 0.7\% & $34\pm 4$ \\
$B\to\eta^\prime\ell\nu$ & 0.5\% & $41\pm 12$ \\ \hline\hline
\end{tabular}
\end{table}

\section{Extraction of Branching Fractions}
\label{sec:extraction_fit_intro}

\subsection{Method and Binning}
\label{sec:binning}

 \begin{table}[tb]
 \begin{center}
 \caption{ \label{tab:PhaseSpaceDivisions} The $q^2$ versus $\cos\theta_{Wl}$ intervals for which we extract partial branching fractions for each decay mode.}
 \label{tab:intervals}
 \begin{tabular}{cccc}\hline\hline
 {\bf Mode}  & {\bf Index} & {\bf $q^2$ Range [GeV$^2$]} & {\bf $\cos\theta_{Wl}$ Range}\\\hline
 $\pi\ell\nu$ & 1 & $0 < q^2 < 2$ & $^-1 < \cos\theta_{Wl} <  1$ \\
$\pi\ell\nu$ & 2 & $2 < q^2 < 8$ & $^-1 < \cos\theta_{Wl} <  1$ \\
$\pi\ell\nu$ & 3 & $8 < q^2 < 16$  & $^-1 < \cos\theta_{Wl} <  1$ \\
$\pi\ell\nu$ & 4 & $q^2 > 16$ & $^-1 < \cos\theta_{Wl} <  1$ \\
 $\rho\ell\nu$ & 1 & $0 < q^2 < 2$ & $^-1 < \cos\theta_{Wl} <  1$\\
 $\rho\ell\nu$ & 2 & $2 < q^2 < 8$ & $^-1 < \cos\theta_{Wl} <  1$\\
 $\rho\ell\nu$ & 3 & $8 < q^2 < 16$ & $0 < \cos\theta_{Wl} <  1$\\
 $\rho\ell\nu$ & 4 & $q^2 > 16$ & $0 < \cos\theta_{Wl} <  1$\\
 $\rho\ell\nu$ & 5 & $q^2 > 8$ & $^-1 < \cos\theta_{Wl} <  0$\\
$\eta\ell\nu$ & - &  all  & $^-1 < \cos\theta_{Wl} <  1$ \\
$\eta^\prime\ell\nu$ & - & 	 all  & $^-1 < \cos\theta_{Wl} <  1$ \\\hline\hline
 \end{tabular}
 \end{center}
 \end{table}

In order to determine partial branching fractions as a function of $q^2$ for $B\to \pi\ell\nu$,
and as a function of both $q^2$ and $\cos\theta_{Wl}$ for $B\to\rho\ell\nu$ we divide
the reconstructed candidates into the regions defined in 
Table~\ref{tab:PhaseSpaceDivisions}.   The variables $\Delta E$ and $M_{h\ell\nu}$ provide sensitivity to signal events and therefore in general we bin the data into seven coarse regions spanning the ranges 
 $5.1750\le M_{h\ell\nu}<5.2875$ GeV/$c^2$ and
$-0.75 <  \Delta E <0.25$ GeV in the $\Delta E-M_{h\ell\nu}$ plane.  We are primarily sensitive to signal within the range defined by $5.2650~\text{GeV}/c^2<M_{h\ell\nu}<5.2875~\text{GeV}/c^2$ and $-0.15~\text{GeV} < \Delta E < 0.25~\text{GeV}$.   The binning choice results from a compromise between sensitivity to the signal processes versus reliance on our simulations to reproduce in detail the signal and background shapes, which depends critically on the modeling of the missing energy and momentum.

In all of the pseudoscalar modes, we obtain  $\Delta E$ versus
$M_{h\ell\nu}$ distributions separately for the $|Q|=1$
and $Q=0$ subsamples.   This separation allows us to take full 
statistical advantage of the cleaner  $Q=0$, while obtaining some
statistical gain from the $|Q|=1$ sample.  Use of the $|Q|=1$ sample
also provides a systematic advantage by reducing our sensitivity to simulation of our
absolute tracking efficiency.

We have limited statistics in the $\eta$ and $\eta'$ modes; therefore, we extract only 
total branching fractions.  However, the backgrounds increase dramatically at large
$q^2$ where the momentum of the  $\eta^{(')}$ is low.  We choose to separate the reconstructed $B\to\eta^\prime\ell\nu$ candidates into two bins: $q^2$ greater and less than 10~GeV$^2$.  Like the division for net charge in the pseudoscalar modes this avoids diluting the purer $q^2<10$~GeV$^2$ sample while still allowing the $q^2>10$~GeV$^2$ sample to contribute to the fit.

Finally, we fit  $\Delta E$ versus $M_{h\ell\nu}$ distributions for each of the
individual $M_{\pi\pi}$ or $M_{3\pi}$ intervals discussed above for modes involving
a $\rho$ or $\omega$, respectively.  The relative yields in these regions provide
resonance line shape information to help separate signal from backgrounds.

To extract the branching fraction information, we fit the reconstructed  $\Delta E$ versus
$M_{h\ell\nu}$ distributions for all modes and all subregions of phase space simultaneously using a binned maximum likelihood procedure.  A summary of the number of bins used in the fit appears in Table~\ref{tab:bin}.

\begin{table}
\caption{Summary of the 532 bins used in the nominal fit. The
different $\rho$ mass bins in the $\eta^{\prime}\to\rho^0\gamma$ 
decay have been categorized as $X_u$ decay bins.} 
\label{tab:bin}
\begin{tabular}{ccccccc}\hline\hline
~ & $M_{h\ell\nu},\Delta E$ & $Q$ & $M_{X_u}$ & $X_u$ Decay & $q^2/\theta_{Wl}$ & {\bf Total}\\ \hline
$\pi^\pm\ell\nu$ & 7 & 2 & 1 & 1 & 4 & {\bf 56} \\
$\pi^0\ell\nu$ & 7 & 2 & 1 & 1 & 4 & {\bf 56} \\
$\rho^\pm\ell\nu$ & 7 & 1 & 3 & 1 & 5 & {\bf 105} \\
$\rho^0\ell\nu$ & 7 & 1 & 3 & 1 & 5 & {\bf 105} \\
$\omega\ell\nu$ & 7 & 1 & 3 & 1 & 2 & {\bf 42} \\
$\eta\ell\nu$ & 7 & 2 & 1 & 2 & 1 & {\bf 28} \\ 
$\eta^\prime\ell\nu$ & 7 & 2 & 1 & 5 & 2 & {\bf 140} \\ \hline\hline
\end{tabular}
\end{table}

The shapes of the $\Delta E$ versus $M_{h\ell\nu}$ distributions for 
signal and background are difficult to parameterize analytically.  
In particular, use of the missing momentum
to estimate the neutrino momentum produces nontrivial 
correlations between $\Delta E$ and $M_{h\ell\nu}$ even though
our reconstruction procedure attempts to minimize correlation.  We therefore rely on MC 
simulations or data studies to provide the components of the fit (described in detail in the following section).  To include the finite statistics of these components, our fitter uses
the method of Barlow and Beeston~\cite{bb:BarlowBeeston}. 

\subsection{Fit Components and Parameters}

We model the contributions of various processes to the data, outlined below, using a combination of MC simulation and independent data samples.  All MC samples incorporate a full GEANT 3~\cite{bb:GEANT} model of the three generations of the CLEO detector, with event samples approximately in the ratio of the number of  $B\bar{B}$ decays from each dataset.  The simulations also track time-dependent detection efficiencies and resolution.

Where applicable, our MC simulation incorporates known inclusive and exclusive $B$ decay modes as of~2004~\cite{elliotprd}.  We correct the generic $B$ model to ensure we can reliably represent background processes and signal efficiency losses due to multiple missing particles, primarily  as a result of $K_L$ particles or charm semileptonic decays within an event.  Based on $K_S\to\pi^+\pi^-$ probes of $K^0$ production, we find that we must increase the weight of events with generated $K_L$ mesons in order to raise the average number of $K_L^0$ per event by a factor of 1.072.   We also corrected the inclusive rate and lepton momentum spectrum of charm semileptonic decays to agree with the convolution of the inclusive $B\to D^{(*)}X$ spectrum~\cite{bdstarspec} with the inclusive charm lepton momentum spectrum from CLEO~\cite{Adam:2006nu}.

Lepton identification efficiency and the rate at which hadrons are misidentified as (``fake'') leptons are difficult to simulate accurately.   In our simulations, we therefore require the reconstructed track for a lepton candidate to originate from a true generator level lepton.  We then apply momentum-dependent electron or muon identification efficiencies obtained from independent data studies.  We then correct for the efficiency loss that arises in our multiple lepton veto due to hadronic fakes.  To do so, we randomly veto events using an event-by-event probability for an event to contain one or more hadronic fakes.  The probability is based on the hadronic content and measured fake rates that are also determined from independent data samples.

We have five categories of components input to our fit.

{\em 1) Signal $B\to X_u\ell\nu$ Decays:}  For our nominal fit, we input signal components using form factors from the unquenched LQCD calculations by HPQCD~\cite{HPQCD_2006} for the $\pi\ell\nu$ modes, form factors from the LCSR results of Ball and Zwicky~\cite{ball04} for the $\rho(\omega)\ell\nu$ modes, and form factors from from the ISGW2 model~\cite{isgw2} for the $\eta\ell\nu$ and $\eta^{\prime}\ell\nu$ modes. 

We divide the signal MC samples at the generator level into the 
$q^2$ and $\cos\theta_{Wl}$ regions summarized in Table~\ref{tab:PhaseSpaceDivisions}.   From each of these subsamples we create the 76 $\Delta E$ versus $M_{h\ell\nu}$ distributions described in Section~\ref{sec:binning} for input to the fit.  Thus we input a generalization of the full efficiency matrix to the fit, and the crossfeed rates among different regions of phase space within a reconstructed mode, as well as among all the different modes, are automatically tied to the observed signal yields.  The sample sizes ranged from 69 to 300 times the expected rates.

Where possible, we use isospin and quark symmetry arguments to combine decay modes in order to increase the statistical precision of the extracted decay rates.  Within $\pi\ell\nu$, the normalization for each of the four $q^2$ regions for  $\pi^\mp\ell^\pm\nu$ floats independently.  The neutral pion rates are constrained to the charged pion rates to be consistent with isospin symmetry, that is $\Gamma(B^0\to\pi^-\ell^+\nu)=2\Gamma(B^+\to\pi^0\ell^+\nu)$.  Similarly the normalization for the five phase space regions in $\rho^\mp\ell^\pm\nu$ float independently, while the five neutral $\rho$ rates are constrained to be consistent with isospin symmetry.  While we reconstruct $\omega\ell\nu$ in only two bins in the $q^2/\cos\theta_{Wl}$ plane, the signal MC is divided into the same five generated regions as the $\rho$ modes.  Based on quark symmetry, we impose the same normalization on the $\rho^0$ and $\omega$ rates.  For our nominal fit we use the 2004 Heavy Flavor Averaging Group~\cite{hfag} values of $f_{+-}/f_{00}=1.026 \pm 0.034$ and $\tau_{B^+}/\tau_{B^0}=1.086\pm0.017$ in applying isospin constraints. 

We fit for the total branching fraction in both the $\eta\ell\nu$ and the $\eta^{\prime}\ell\nu$ modes.  The branching fractions for each $\eta$ and $\eta^\prime$ decay mode are fixed to be consistent with~\cite{pdg06}.  In $\eta^{\prime}\ell\nu$, the relative signal strengths in the two $q^2$ bins are fixed by the rate predictions of the input form factor. 

In total, the partial rates in each of the phase space bins for $B^0\to\pi^-\ell^+\nu$ and $B^0\to\rho^-\ell^+\nu$ in addition to the total $B^+\to\eta\ell^+\nu$ and $B^+\to\eta^\prime\ell^+\nu$ are described by eleven free parameters.

{\em 2) $b\to c$ Decays:} The dominant $b\to c$ background is modeled using a Monte-Carlo simulation that incorporates known inclusive and exclusive $B$ branching fractions.  To reduce systematic sensitivity in simulation of this component, we allow the  $b\to c$ normalization to float independently for each of the reconstructed modes, net charge bins, and hadronic submodes.  This procedure introduces fifteen free parameters for a total of 26 free parameters in the fit.  

{\em 3) Continuum Background:} The fit components for the $e^+e^-\to q\bar{q}$ background are
obtained using data samples collected 60 MeV$/c^2$ below the $\Upsilon(4S)$ peak.  These samples have luminosities of $1/2$ ($1/3$) the total CLEO~II/II.5 (CLEO~III) on-resonance ($B\bar{B}$) luminosity.  While $q\bar{q}$ processes dominate this sample, it also accounts for residual two-photon and $\tau^+\tau^-$ backgrounds.  In reconstruction, we account for the shift in energy by scaling $M_{h\ell\nu}$ by $E_{\text{beam}}^{\text{on}}/E_{\text{beam}}^{\text{off}}$ and, for the neutrino rotation in the $q^2$ calculation, scaling $M_B$ by the inverse energy ratio.

The combination of our continuum suppression and full $B$ reconstruction requirements results in low continuum backgrounds overall.  As a result,  the off-resonance distributions have many bins in the Poisson statistics  regime.  Because the off-resonance data must be scaled upward to match the  on-resonance luminosity,  the Barlow-Beeston extension of the likelihood procedure can still result in bias from the very low statistics bins.   

\begin{figure}
\begin{center}
\includegraphics[width=2.5in]{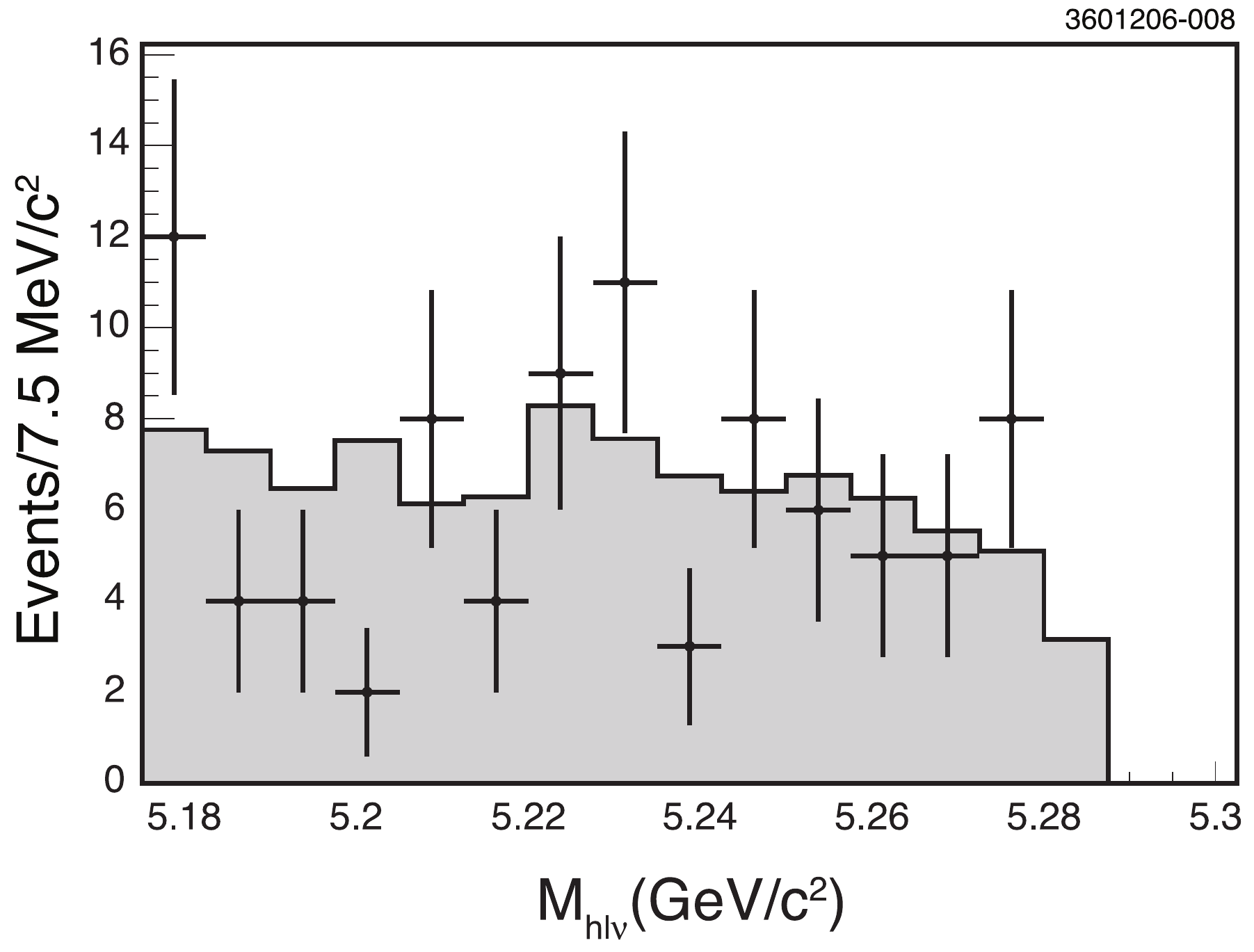}
\caption{\label{fig:contsmooth}The $M_{h\ell\nu}$ distribution for $B\to\pi\ell\nu$ candidates from off resonance data before (points with errors) and after (filled histogram) the continuum smoothing procedure.}
\end{center}
\end{figure}

To avoid bias, we adopt a continuum smoothing procedure in which we reanalyze the off-resonance data with the explicit continuum suppression criteria removed.   This procedure produces a relatively high-statistics set of $M_{h\ell\nu}-\Delta E$  distributions.   We correct each distribution for potential bias caused by the removal of the continuum suppression.  The correction is obtained from comparing analysis of an independent mock continuum sample with and without the continuum suppression criteria of our analysis.  MC provides the $e^+e^-\to q\bar{q},\,\tau^+\tau^-$ component with true (generator-level) reconstructed leptons.  The fake lepton component is obtained from sample of purely hadronic off-resonance data events combined with measured fake rates.  After application of the bias correction,  each distribution is normalized so that the integrated rate in the 
$M_{h\ell\nu}-\Delta E$ region input to the fit is identical to the integrated rate obtained from the off-resonance data with the continuum suppression.  Figure~\ref{fig:contsmooth} shows one example.  

The off-resonance component is absolutely normalized in the fit based on the ratio of on- to off-resonance integrated luminosity  with a correction  for the energy dependence of the $e^+e^-\to q\bar{q}$ cross section.

{\em 4) $B\bar{B}$ Fake Signal Leptons: }  Both $B\bar{B}$ and continuum processes contribute backgrounds where a hadron has faked a lepton.  Our continuum sample incorporates this contribution, but, as noted above, we do not rely on MC for the background component resulting from fake leptons. 

We obtain the $B\bar{B}$ fake component from analysis of on-resonance and off-resonance data samples in which no leptons have been identified.  For each reconstructed track satisfying our electron kinematic and fiducial 
criteria in each of these events, we analyze the event treating that track as the signal electron.  If the event with that fake lepton candidate satisfies our analysis, then a contribution 
weighted by the measured electron fake rate for a track of that momentum is added.
Each track is then similarly
treated as a muon candidate.  The on- and off-resonance samples are absolutely normalized in the fit based on luminosity.  The off-resonance receives a negative normalization, causing the continuum component to be subtracted from the on-resonance data, thereby isolating the $B\bar{B}$ component.

{\em 5) Other $B\to X_u\ell\nu$ Decays:}  A final background arises from other $B\to X_u\ell\nu$ decays that we are not exclusively reconstructing in the fit.  To model this background we use a hybrid exclusive-inclusive MC developed and documented by Meyer~\cite{tomthesis} that combines ISGW2 predictions of exclusive decays~\cite{isgw2} with the inclusive lepton spectrum predicted using HQET by De Fazio and Neubert~\cite{defazio}.  The parameters used in the heavy quark expansion are constrained by measurements of the $B\to X_s\gamma$ photon spectrum by CLEO~\cite{cleobsgprl}.  The model generates resonant decays according to ISGW2 plus non-resonant decays such that the total generated spectrum matches theoretical predictions.  We explicitly remove our exclusive decays from this simulation, and the remainder of the sample becomes the $B\to X_u^{\text{other}}\ell\nu$ component of the fit.

The normalization of this component uses a measurement of the inclusive charmless leptonic  branching fraction by the \textsc{BaBar} collaboration~\cite{babarinclep} in the momentum endpoint range of 2.2 to 2.6 GeV$/c$.  \textsc{BaBar}  finds $\mathcal{B}(B\to X_u e^+\nu)_{\text{endpoint}}=(2.35\pm0.22)\times 10^{-4}$.   Roughly 10\% of the generated $B\to X_u^{\text{other}}\ell\nu$ spectrum has a lepton in this range.   For each iteration of the fit, we recalculate the normalization of the ``Other $B\to X_u\ell\nu$'' component so that the contributions of this component and of the seven signal modes to the endpoint region sum to the measured rate.

\subsection{Verification of Fitting Procedure}
To test the fitter for systematic biases, we performed a
bootstrap study using 214 mock data samples which were generated by randomly selecting
subsets of our MC events.
For each mock sample, the numbers of events
for each signal and background contribution were fixed to our
efficiency--corrected yields. From fits to the 214 analyzed subsets, we find that the fitter
reproduces the branching fractions without bias within the statistical uncertainties of the procedure (about 7\% of the reported statistical error on a given quantity).  Furthermore, the parametric uncertainties reported for the fits agree well with the widths of the distributions of the parameters' central values.   

The fit implementation was also cross-checked against an independent implementation used in the previous analysis~\cite{bb:Athar:2003yg} on that analysis' data.

\begin{table}
\caption{\label{tab:parsum}A summary of the different components and the corresponding fill style in the subsequent histograms.}
\begin{tabular}{clcl}
\hline\hline
\parbox[c]{0.4in}{\includegraphics{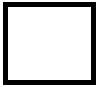}} & signal & \parbox[c]{0.4in}{\includegraphics{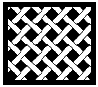}} & $B\to X_u\ell\nu$ other  \\
 \parbox[c]{0.4in}{\includegraphics{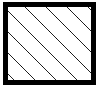}} & $\pi$ cross-feed & \parbox[c]{0.4in}{\includegraphics{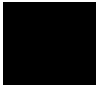}} & fake lepton   \\
\parbox[c]{0.4in}{\includegraphics{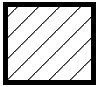}} & $\rho/\omega$ cross-feed &   \parbox[c]{0.4in}{\includegraphics{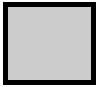}} & continuum \\
\parbox[c]{0.4in}{\includegraphics{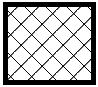}} & $\eta/\eta^{\prime}$ cross-feed & \parbox[c]{0.4in}{\includegraphics{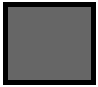}} & $b\to c$   \\ \hline \hline
\end{tabular}
\end{table}

\subsection{Fit Results}
\label{sec:fit_results}

\begin{figure*}[tp]
\begin{center}
\includegraphics[width=\linewidth]{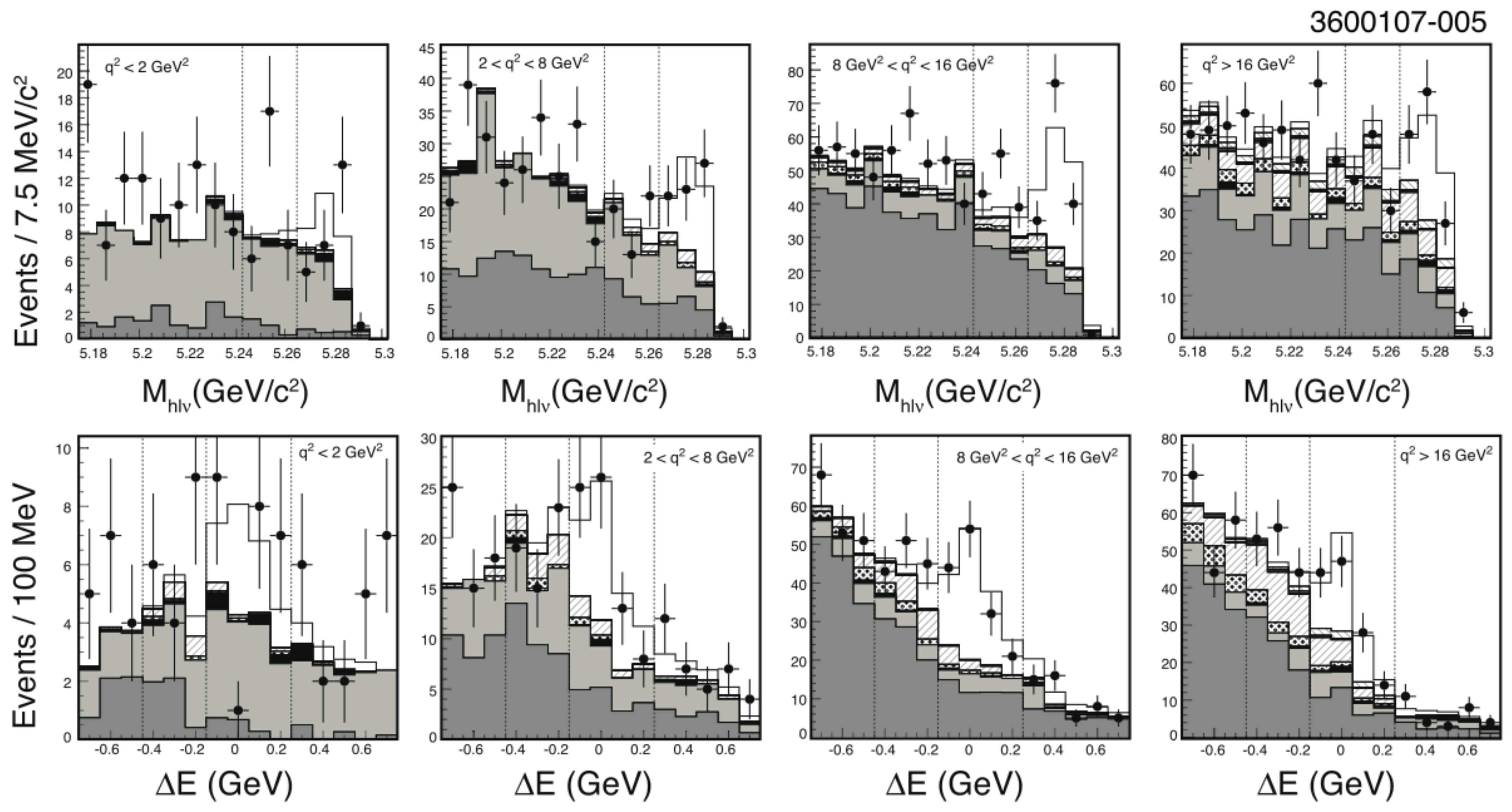}
\caption{\label{fig:firstfig} The $M_{h\ell\nu}$ and $\Delta E$ projections of the nominal fit, $Q=0$, for the summed $\pi^\pm\ell\nu$ and $\pi^0\ell\nu$ modes.  The fit components are described in Table~\protect\ref{tab:parsum}.}
\end{center}
\end{figure*}

\begin{figure*}[tb]
\begin{center}
\includegraphics[width=\linewidth]{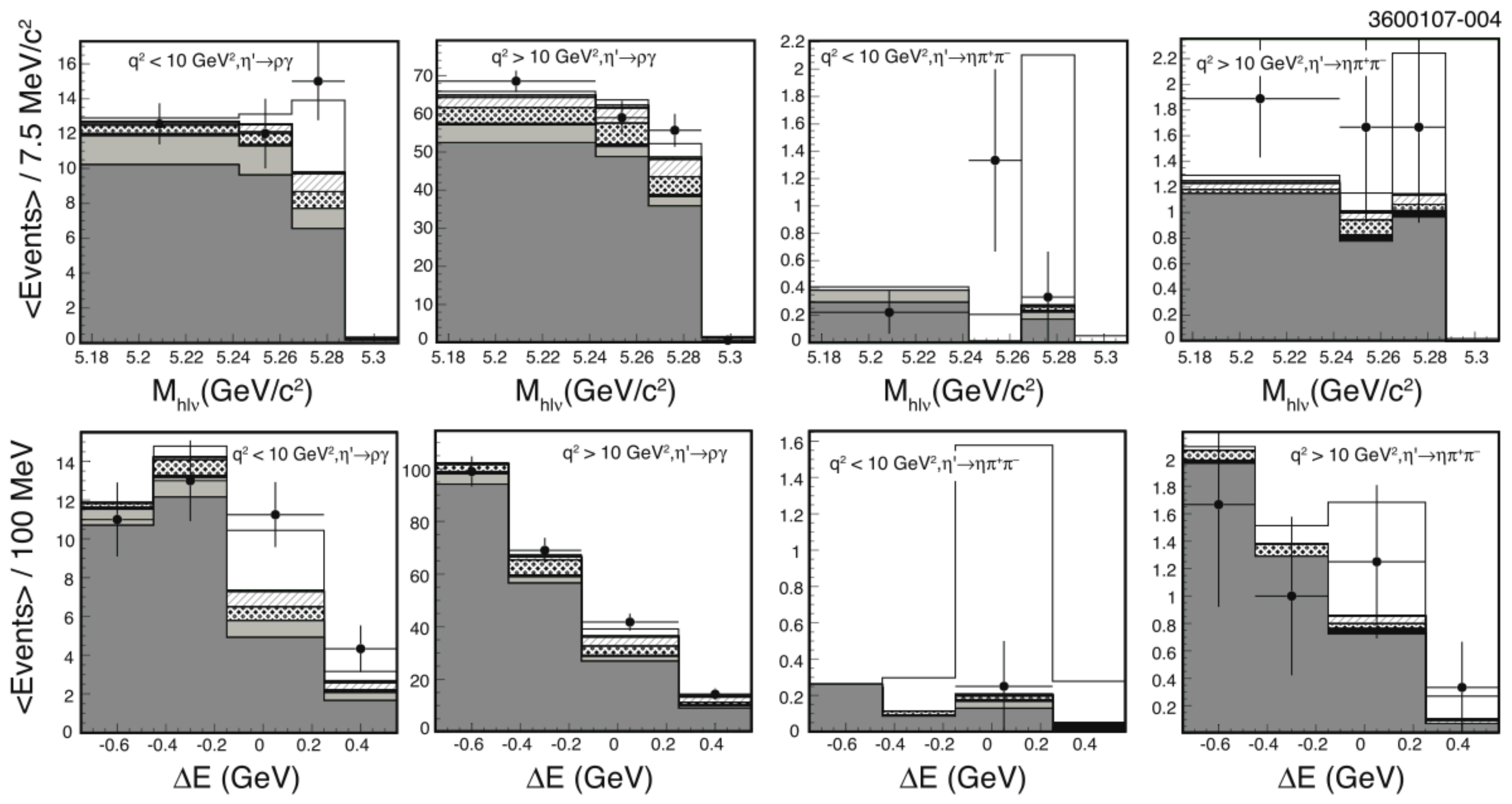}
\caption{\label{fig:lastfig} The $M_{h\ell\nu}$ and $\Delta E$ projections of the nominal fit, $|Q|=0$, for the $\eta^\prime\ell\nu$ mode.  The fit components are described in Table~\protect\ref{tab:parsum}.  The $\eta^\prime\to\rho\gamma$ and $\eta^\prime\to\eta\pi\pi$ submodes have been summed.}
\end{center}
\end{figure*}

Table ~\ref{tab:results} summarizes the branching fractions obtained in
the nominal fit for the $\pi\ell\nu$, $\rho\ell\nu$, $\eta\ell\nu$, and
$\eta^{\prime}\ell\nu$ modes.  The nominal fit converges with $-2\ln\mathcal{L}$ equal to 541 and we note that the statistical errors on some of the 532 bins are not Gaussian.  Figures~\ref{fig:firstfig} -~\ref{fig:lastfig} show $M_{h\ell\nu}$ and $\Delta E$ projections for the nominal fit.  These figures are generated by plotting $M_{h\ell\nu}$ for candidates with $\Delta E$ in the signal bin and vice versa.  Note that for the $\pi$ and $\rho$ modes, the binning is finer than that which the fitter uses in order to show detailed peak structure.  Raw signal yields for the various decay modes, integrated over phase space, are listed in Table~\ref{tab:eff}.

\begin{figure*}[tp]
\begin{center}
\includegraphics[width=5.6in]{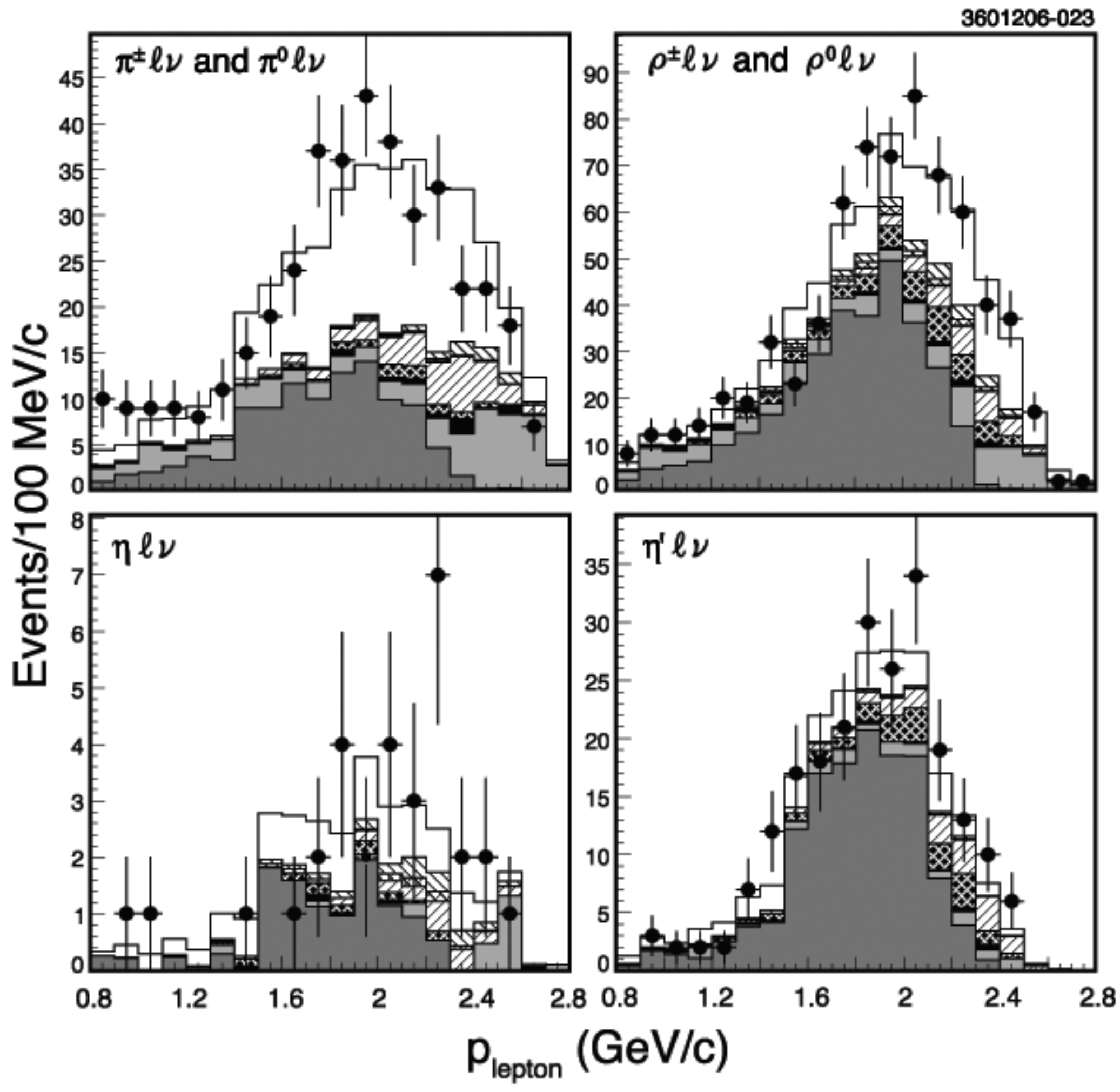}
\caption{\label{fig:lepmomproj}Lepton momentum projections of the nominal fit for $\Delta Q=0$, $\pi^\pm\ell\nu$ and $\pi^0\ell\nu$ (top left), $\rho^\pm\ell\nu$ and $\rho^0\ell\nu$ (top right), $\eta\ell\nu$ (bottom left) and $\eta^\prime\ell\nu$ (bottom right).  The data have been summed over $q^2$, $\cos\theta_{Wl}$, and, where applicable, decay mode.}
\end{center}
\end{figure*}

Figure~\ref{fig:lepmomproj} shows the lepton momentum spectrum for $B\to\pi\ell\nu$ and $B\to\rho\ell\nu$ events in the signal $\Delta E$--$M_{h\ell\nu}$ bin.  As expected, the signal lepton spectrum for $B\to\rho\ell\nu$ is noticeably harder than that for $B\to\pi\ell\nu$.  The agreement between the data and the sum of the fit components gives us confidence in the overall fit quality and our in ability to model the lepton momentum distribution of the fit components.

\begin{figure*}[tbp]
\begin{center}
\includegraphics[width=5.6in]{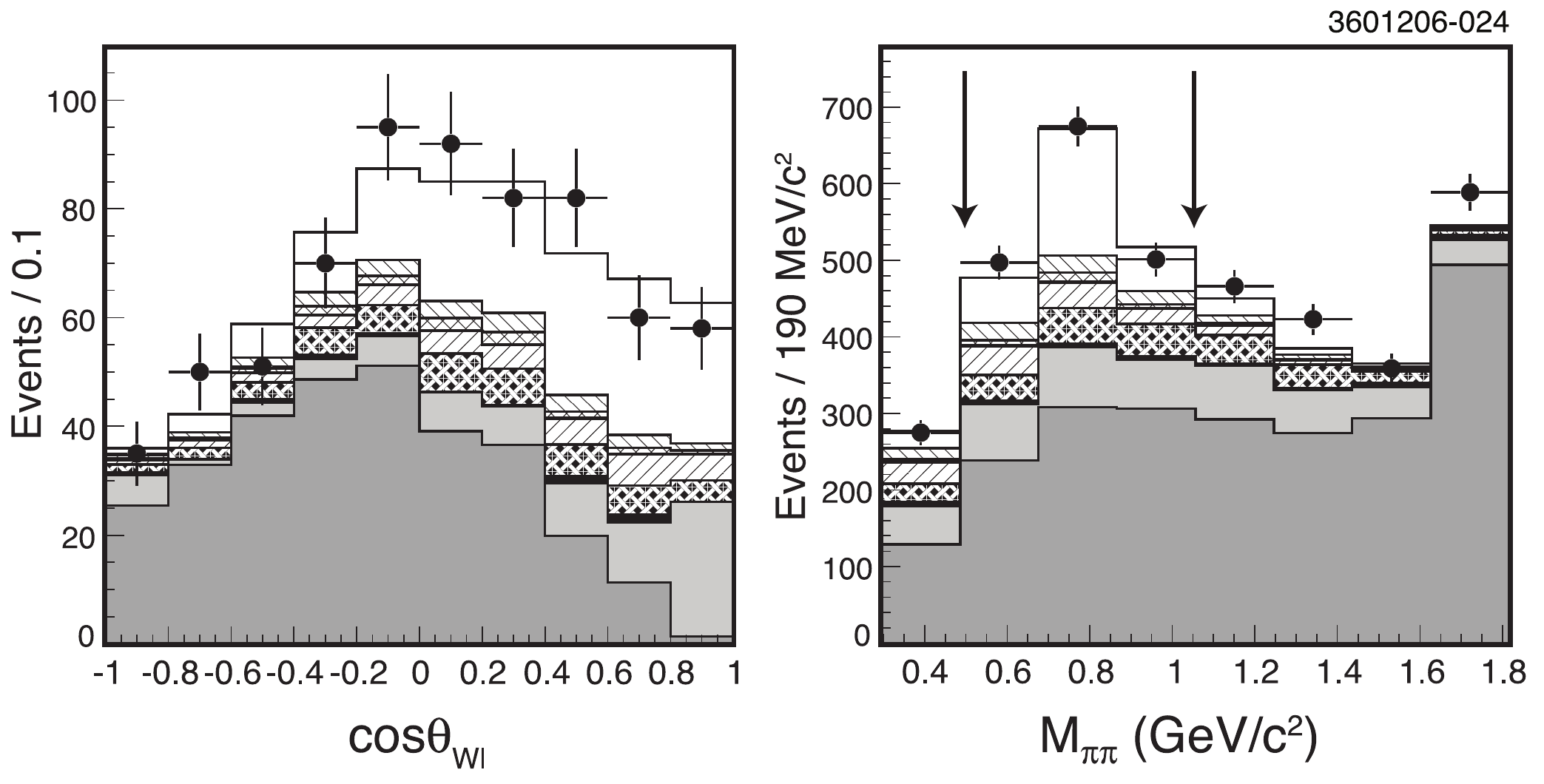}
\caption{\label{fig:rhoppmproj} The $\cos\theta_{Wl}$ projection of the nominal fit, summed over $q^2$, for the $\rho^0\ell\nu$ and $\rho^\pm\ell\nu$ modes (left).  The projection of the two-pion invariant mass of the $\rho$ candidates summed over $q^2$ and $\cos\theta_{Wl}$ (right).  Arrows indicate the region included in the fit.  The peak at right is due to $B\to D^{(*)}\ell\nu$, with $D\to K\pi$ and the $K$ misidentified as a $\pi$.}
\end{center}
\end{figure*}

Figure~\ref{fig:rhoppmproj} shows the projections of $\cos\theta_{Wl}$ and the invariant mass of the two pions, $M_{\pi\pi}$, used to construct the $\rho$ candidate for the $\rho\ell\nu$ modes.  Again, the agreement between data and the sum of the fit components is excellent.  In the $\cos\theta_{Wl}$ distribution the signal accumulates near one and the $b\to c$ background is concentrated in the area where $\cos\theta_{Wl}$ is less than zero as expected.  The $M_{\pi\pi}$ projection shows a clear resonance component and extrapolates reliably into regions not used in the fit.  The peak in the highest bin is due to $B\to D^{(*)}\ell\nu$ where the $D\to K\pi$ decay is identified as a $\rho\to\pi\pi$ decay.  

\begin{figure*}[tp]
\begin{center}
\includegraphics[width=\linewidth]{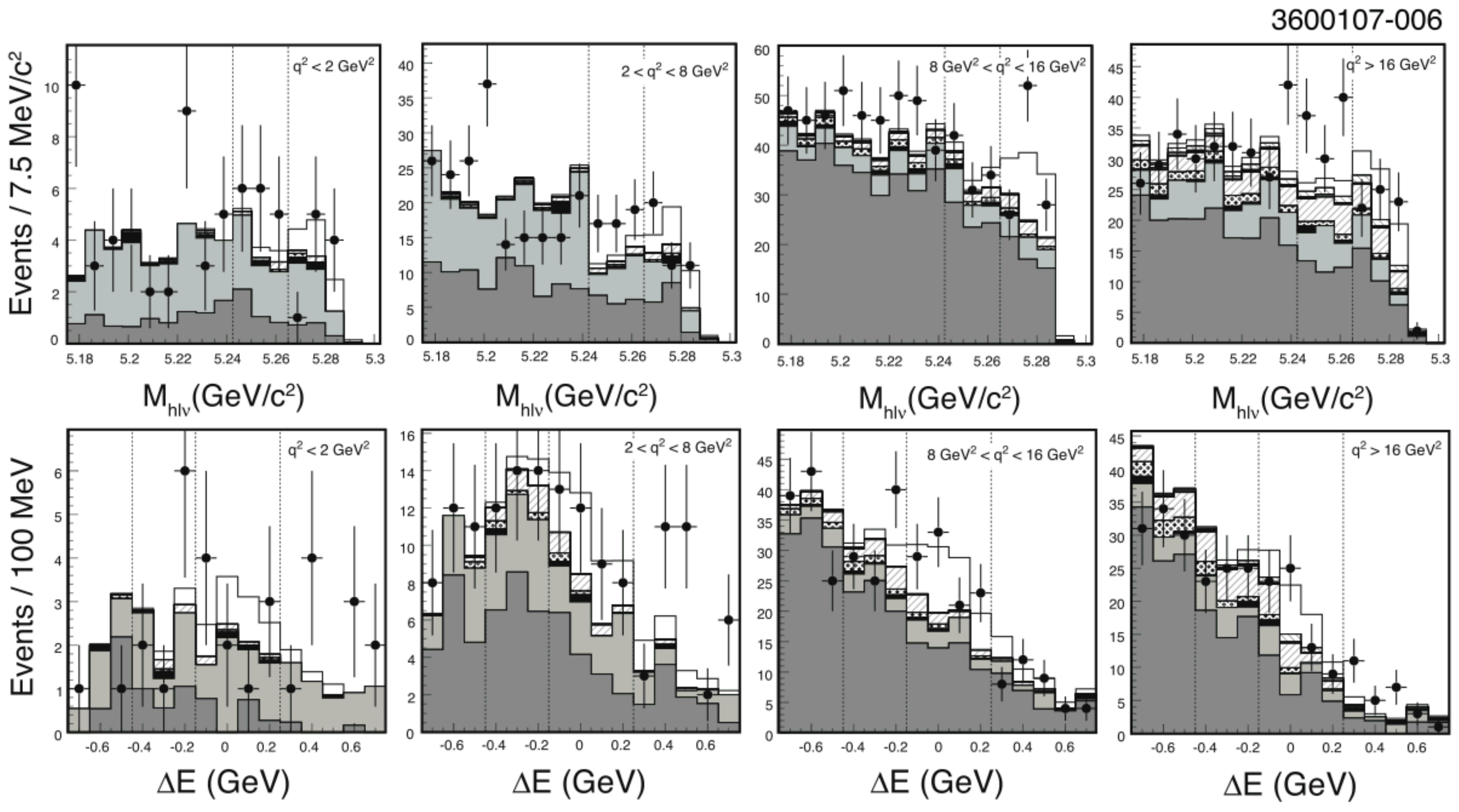}
\caption{The $M_{h\ell\nu}$ and $\Delta E$ projections of the nominal fit, $|Q|=1$, for the summed $\pi^\pm\ell\nu$ and $\pi^0\ell\nu$ modes.  The fit components are described in Table~\protect\ref{tab:parsum}.}
\end{center}
\end{figure*}

\begin{figure*}[p]
\begin{center}
\includegraphics[width=\linewidth]{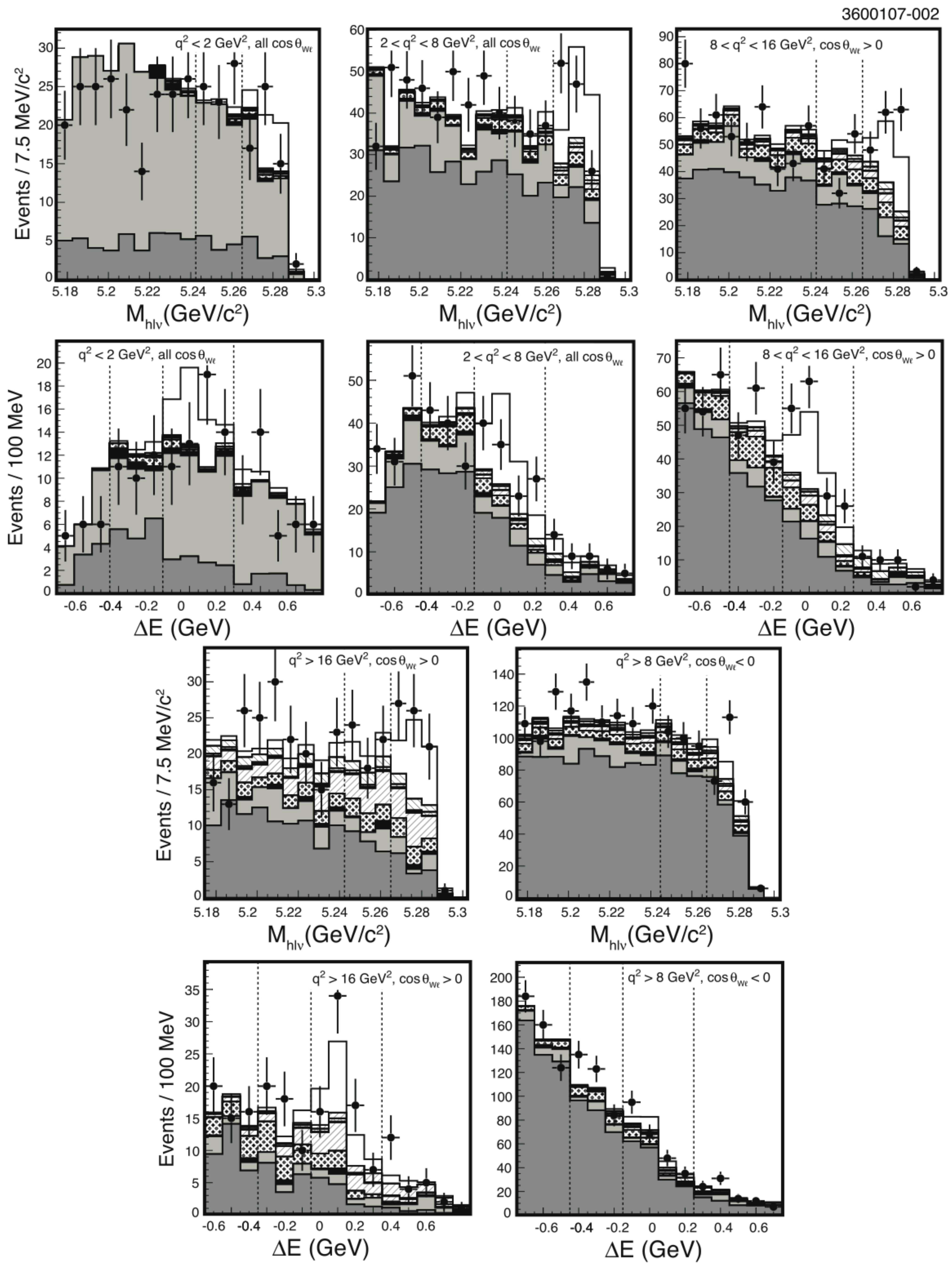}
\caption{The $M_{h\ell\nu}$ and $\Delta E$ projections of the nominal fit for the summed $\rho^\pm\ell\nu$ and $\rho^0\ell\nu$ modes.  The fit components are described in Table~\protect\ref{tab:parsum}.}
\end{center}
\end{figure*}

\begin{figure*}[tp]
\begin{center}
\includegraphics[width=\linewidth]{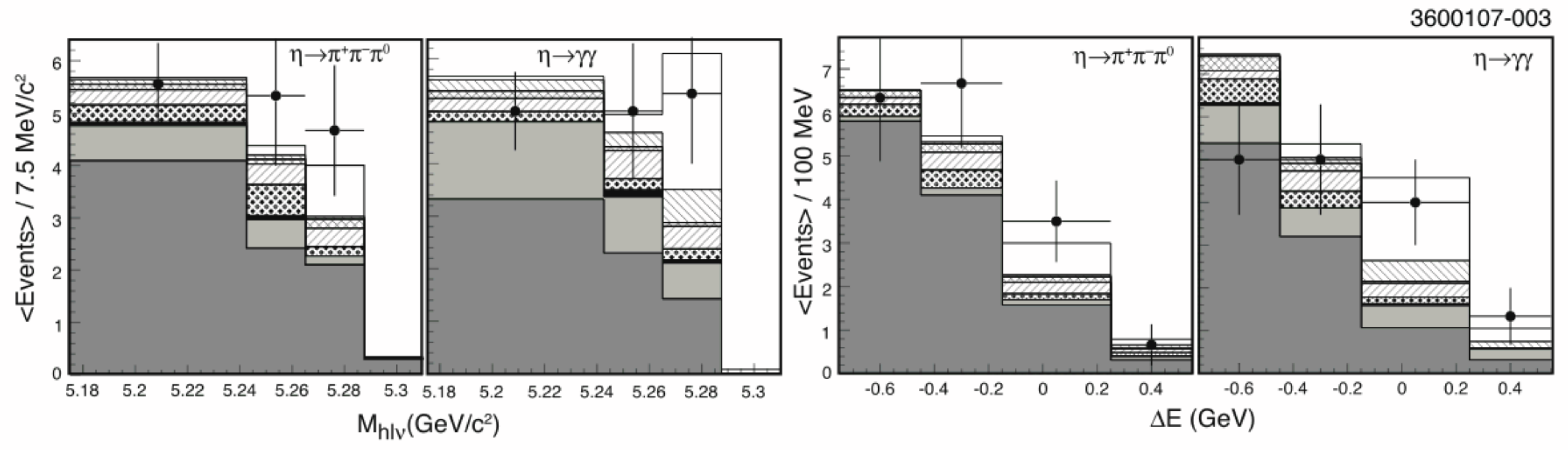}
\caption{The $M_{h\ell\nu}$ and $\Delta E$ projections of the nominal fit, $|Q|=0$, for the $\eta\ell\nu$ mode.  The fit components are described in Table~\protect\ref{tab:parsum}.}
\end{center}
\end{figure*}

\begin{table}
\caption{Summary of the phase space subregions and the partial and total branching fraction results.  The errors shown are statistical and systematic, respectively.}
\label{tab:results}
\begin{tabular}{cccc}\hline\hline
~ & $q^2$ [GeV$^2$]    & $\cos\theta_{Wl}$ & $\mathcal{B}$ [$10^{-4}$] \\ \hline
$B^0\to\pi^-\ell^+\nu$   & 0 - 2   & $^-1$ - 1  & $0.13\pm 0.07 \pm 0.02$ \\ 
~                           & 2 - 8   & $^-1$ - 1 & $0.27\pm 0.08 \pm 0.03$ \\
~                           & 8 - 16 & $^-1$ - 1 & $0.56\pm 0.09 \pm 0.05$ \\
~                           &$>16$ & $^-1$ - 1 & $0.41\pm 0.08 \pm 0.04$ \\ 
~                           & \multicolumn{2}{c}{all phase space} & ${\bf 1.37 \pm 0.15 \pm 0.11}$ \\
$B^0\to\rho^-\ell^+\nu$ & 0 - 2   & $^-1$ - 1 & $0.45\pm 0.20\pm 0.15$ \\
~                             & 2 - 8   & $^-1$ - 1 & $0.96\pm 0.20\pm 0.29$ \\
~                             & 8 - 16 &  0 - 1 & $0.75\pm 0.16\pm 0.14$ \\
~                             & $>16$    &  0 - 1 & $0.35\pm 0.07\pm 0.05$ \\
~                             & $>8$     & $^-1$ - 0 & $0.42 \pm 0.18 \pm 0.31$ \\ 
~                             & \multicolumn{2}{c}{all phase space} & ${\bf 2.93 \pm 0.37 \pm 0.37}$ \\
$B^0\to\eta\ell^+\nu$ & \multicolumn{2}{c}{all phase space} & ${\bf 0.44 \pm 0.23 \pm 0.11}$ \\ 
$B^0\to\eta^\prime\ell^+\nu$ & \multicolumn{2}{c}{all phase space} & ${\bf 2.66 \pm 0.80 \pm  0.56}$ \\ \hline \hline
\end{tabular}
\end{table}

\section{Systematic Uncertainties}
\label{sec:systematics}

The systematic uncertainty in this analysis are dominated by effects that couple to accurate simulation of the missing momentum and energy used to reconstruct the neutrino, and to simulation of the selection criteria to isolate events where this association works well.  Additional uncertainties enter from modeling of the background processes.  We explicitly evaluate our residual uncertainty due to form factor shapes.

For most systematic effects, such as simulation of the absolute tracking efficiency, we have information from independent studies that limit their size.  For most studies, statistics limit the sensitivity, so a Gaussian treatment is not unreasonable.  To assess the effect of a given systematic on this analysis, we completely recreate the fit components with the suspect quantity biased to the limit of the study.  This typically involves reweighting or biasing of the generated MC samples, which minimizes statistical issues in the determination of the shifts.  For studies involving use of random numbers ({\it e.g.}, discarding tracks to decrease the tracking efficiency), we repeat the study several times with different seeds to ensure that we have an accurate measure.   We then refit, and assign the difference from the nominal fit as the systematic estimate.   Note that we apply the bias to the same samples used to create the nominal fit components, which largely eliminates statistical fluctuations in the procedure.

This procedure will, in principle, cause compensating changes between the efficiencies input to the fit and the signal yields before efficiency correction obtained from the fit.  Degrading resolution, for example, mainly smears signal out of the signal region (decreased efficiency), but allows greater latitude for backgrounds to smear into the signal region (increased background, so decreased signal yield). Since the systematic uncertainties can result in improper estimation of this
cancellation, we will trust any cancellation to only 30\% of itself, adding that portion back into the assessed uncertainty.

 Because of correlations between yields in different regions of phase space, the systematic uncertainty for the total branching fraction can be smaller than uncertainties for the individual partial branching
fractions.  

 \subsection{Systematic Uncertainties in Neutrino Reconstruction}
 
The  systematic uncertainties associated with neutrino reconstruction efficiency and resolution, summarized in Table~\ref{tab:nureconsyst}, dominate the systematic uncertainties in this analysis. 

{\em Track Reconstruction Efficiency:}  Track reconstruction efficiency agrees  in data and simulation to within 
2.6\% (5\%) of itself for low momentum tracks (under 250 MeV$/c$) for CLEO~II/II.5 (III) and 
to within 0.5\% for high momentum tracks.  We randomly discard tracks at these levels to assess the systematic effect.  We conservatively assume that the uncertainties are fully correlated among the three detector configurations.  

{\em Track Momentum Resolution:}  Based on studies of the
reconstructed $K\pi$ mass resolution in $D\to K\pi$ at the
$\Upsilon(4S)$, we increase the deviations between reconstructed
and generated track momenta by 10\% (40\%) in CLEO~II/II.5 (III).
The CLEO~II/II.5 value is very conservative, while CLEO~III value
brings data and simulation into proper agreement.

{\em Shower Reconstruction Efficiency:}  Shower reconstruction efficiency in data and simulation agree within 1.6\% (1\%) of itself for CLEO~II/II.5 (III).   We therefore randomly discard a fraction of the showers consistent with these limits.  We
again assume full correlation among different detector generations
when modifying the simulation.

{\em Shower Resolution:}  
We increase all photon mismeasurements by 10\% of themselves.
 
{\em Hadronic Showering ``Splitoff'' Simulation and Rejection:}
We find that the number of splitoff clusters reconstructed from hadronic
showers in our simulation deviate from the number in data by at most
0.03 per hadron.   The energy spectrum agrees well.  We therefore add additional 
showers with the observed spectrum at the rate of 0.03 per hadron in each event.

We also systematically distort the neural-net rejection
variable, which is based upon shower shapes, until the data and simulation
clearly disagree.  We do this for both showers originating from photons and from
hadronic shower simulations.
  
{\em Particle Identification:}   Particle identification uncertainties directly affect the missing 
energy resolution.  The device-based particle identification efficiencies 
are understood at the level of 5\% or better, and our uncertainties as
a function of detector and momentum are considered in our
systematic evaluation.  The overall effect of these uncertainties
gets reduced by our weighting according to particle production
probability.

{\em $K^0_L$ Production and Energy Deposition:} 
Our $K_L$ production rate correction is modified according to the
statistical uncertainties of our $K_S$ data versus simulation study.
A separate study of kaons that shower in the CsI calorimeter indicate
that the deposited energy is simulated to within 20\%.

{\em Secondary Lepton Spectrum:}
We vary corrections based on the measurements of the 
inclusive $B\to D^{(*)}X$ spectrum and the electron momentum spectrum in 
inclusive semileptonic charm decay according to the envelope of
potential $b\to c\to s\ell\nu$ lepton spectra resulting from the uncertainties
in the spectral measurements.
 
\begin{sidewaystable}
 \begin{center}
 \caption{ \label{tab:nureconsyst}The systematic errors associated with neutrino reconstruction.   Errors from different sources are assumed to be uncorrelated.  The bottom row shows the quadrature-sum of the errors.  The phase space interval indices are as defined in Table~\ref{tab:intervals}.}
  \begin{tabular}{lcccccccccccccc}\hline\hline
  & \multicolumn{5}{c}{$\delta\mathcal{B}\left(B^0\to\pi^-\ell^+\nu\right)~[\%]$} & &  \multicolumn{6}{c}{$\delta\mathcal{B}\left(B^0\to\rho^-\ell^+\nu\right)~[\%]$} & \multicolumn{2}{c}{$\delta\mathcal{B}\left(B^+\to(\eta\slash\eta^{\prime})\ell^+\nu\right)~[\%]$} \\
  \multicolumn{1}{c}{ {\bf Systematic Error Source} }& 1 & 2 & 3 & 4 & All & ~~~ & 1 & 2 & 3 & 4 & 5 & All & \hspace{25pt}$\eta$\hspace{25pt} & $\eta^{\prime}$ \\\hline
  Track Efficiency & 2.8 & 3.4 & 4.0 & 2.4 & 2.5 & & 8.4 & 7.9 & 11.9 & 7.7 & 16.0 & 5.5 & 3.9 & 6.1 \\
  Track Resolution & 1.7 & 2.1 & 1.8 & 2.7 & 1.4 & & 10.3 & 16.2 & 0.8 & 2.8 & 50.0 & 1.0 & 6.3 & 9.6 \\
  Shower Efficiency & 5.5 & 3.9 & 3.6 & 3.0 & 3.1 & &  7.0 & 8.7 & 7.0 & 5.9 & 38.1 & 3.2 & 8.5 & 8.4 \\
  Shower Resolution & 2.2 & 4.4 & 3.3 & 2.3 & 3.1 & &  8.3 & 7.0 & 7.2 & 2.8 & 4.0 & 5.2 & 19.6 & 10.3 \\
  Splitoff Simulation & 3.1 & 2.1 & 2.6 & 2.6 & 2.2 & &  6.7 & 8.3 & 3.1 & 3.4 & 22.1 & 1.4 & 4.1 & 3.4 \\
  Splitoff Rejection & 0.0 & 0.0 & 0.1 & 0.2 & 0.1 & & 0.2  & 0.1 & 0.1 & 0.0 & 0.7 & 0.1 & 0.0 & 5.6 \\
  Particle Identification & 0.7 & 2.2 & 0.9 & 4.4 & 1.1 & &  7.5 & 12.7 & 0.8 & 3.4 & 18.0 & 2.4 & 2.5 & 3.5 \\
  $K^0_L$ Production  & 0.2 & 0.1 & 0.2 & 0.2 & 0.2 & & 0.5 & 0.9 & 0.1 & 0.3 & 3.6 & 0.1 & 0.1 & 0.1 \\
  $K^0_L$ Energy Deposition & 0.0 & 0.0 & 0.2 & 0.3 &  0.2 & & 1.9 & 2.2 & 0.6 & 1.1 & 8.0 & 0.9 & 3.7 & 3.3 \\
  Secondary Lepton Spectrum & 0.5 & 1.1 & 0.4 & 0.4 & 0.2 & & 6.7 & 0.6 & 0.6 & 0.2 & 2.8 & 1.5 & 0.3 & 0.3 \\ 
  \parbox[c]{1.5in}{ \vspace{0.1in} {\bf Sum } \vspace{0.1in} } & {\bf 7.5} & {\bf 7.9} & {\bf 7.1} & {\bf 7.4} & {\bf 5.9} & & {\bf 21.1} & {\bf 26.2} & {\bf 16.0} & {\bf 11.6} & {\bf 71.5} & {\bf 8.9} & {\bf 23.4} & {\bf 19.3} \\
  \hline\hline
  \end{tabular}
  \end{center}
  \end{sidewaystable}
  
  \subsection{Additional Sources of Systematic Error}

The full list of systematic uncertainties are summarized
in Table~\ref{tab:expsyst} and described below.

{\em Continuum Suppression:} We vary the independent parameters in the functions used in the continuum smoothing algorithm that parameterize the efficiency of our continuum suppression algorithm in the $M_{h\ell\nu}-\Delta E$ plane within their uncertainties. 

{\em $b\to c$:} We vary the  $B\to D\ell\nu$, $B\to D^*\ell\nu$, 
$B\to D^{**}\ell\nu$, and non-resonant $B\to X_c\ell\nu$ branching fractions
according to the uncertainties obtained in Reference~\cite{elliotprd}.   We also 
apply identical variations to the $B\to D^*\ell\nu$ 
form factors that are outlined in Reference~\cite{elliotthesis}. 

{\em Other $B \to X_u\ell\nu$:} We vary the heavy quark expansion (HQE) parameters~\cite{defazio} at the heart of our
hybrid $B \to X_u\ell\nu$ generator consistent with uncertainties in the CLEO $B\to X_s\gamma$ 
photon spectrum measurement~\cite{cleobsgprl}, which is conservative
compared to current world knowledge.  We also probe the hadronization
uncertainty by changing the rate for the exclusive modes (not including our signal modes) by  
$\pm10\%$, simultaneously adjusting the rate of the inclusively generated portion to maintain the same total rate.

We vary the endpoint rate to which our $B\to X_u\ell\nu$ component is fixed, within the uncertainties of the \textsc{BaBar} endpoint measurement~\cite{babarinclep}.  All uncertainties are combined in the summary table, though
the HQE and endpoint constraint uncertainties dominate.

{\em Lepton Identification and Fake Leptons:}
We have measured our lepton identification efficiencies  at
the 2\% level.  We also vary the rate at which hadrons fake leptons by $\pm 50\%$.  In momentum regions with
measurable fake rates, this variation is very conservative.

{\em $\pi^0$ Identification:} 
The CLEO~III MC simulation overestimates the $\pi^0$ efficiency by $4\pm2\%$.  
Therefore we apply a correction of 0.96 to the CLEO~III MC 
signal decays that contain a  $\pi^0$ in the final state.  Note that $\pi^0$ finding is only used in the 
process of reconstructing signal decays. To assess the systematic 
uncertainty,  we remove this efficiency correction in CLEO~III.

{\em Number of $\Upsilon\to B\bar{B}$ Events:}
Based on luminosity, cross section, and event shape studies the error on the 
number of $B\bar{B}$ events in CLEO~II~+~II.5~(III) is taken to be 2\% (8\%).  
Combining these uncertainties and accounting for
the relative luminosities in each, we find an uncertainty on the total
number of $B\bar{B}$ events, and therefore the branching fractions, to be 3.6\%.

{\em $\tau_{B^+}/\tau_{B^0}$ and $f_{+-}/f_{00}$:}
The fixed relative strengths of the charged and neutral rates are sensitive to 
both the lifetime and production ratios of charged and neutral $B$ mesons.
Since the fit constraints on the charged and neutral $\pi$ and $\rho$ depend on 
$f_{+-}/f_{00}$, this systematic affects both the total yields from the fit and the conversion
factor from the yields to branching fractions.   
We vary both the lifetime and production fraction ratios independently 
within the uncertainties produced by the Heavy Flavor Averaging Group~\cite{hfag}.

{\em Final State Radiation:}
Final state radiation corrections calculated using PHOTOS~\cite{photos} are applied to
 the nominal results. We assign an uncertainty by varying final state radiation corrections by 20\%. 

{\em Non-Resonant $B\to\pi\pi\ell\nu$:}
In the $B\to\rho\ell\nu$ modes there is an additional uncertainty from the unknown 
contribution of non-resonant $B\to\pi\pi\ell\nu$.  Using isospin and angular momentum arguments outlined in~\cite{bb:Athar:2003yg}, we expect the dominant non-resonant $\pi\pi$ contribution to be in $\pi^+\pi^-$ and $\pi^0\pi^0$.  While extracting a true non-resonant  $\pi\pi\ell\nu$ yield is difficult with some knowledge of the form factors and mass distribution, we can limit the amount that a $\rho$~line~shape can be projected out of the $\pi^0\pi^0\ell\nu$ data, and use the isospin relationships to, in turn, bound the non-resonant component in $\pi^+\pi^-\ell\nu$.  Procedurally, we add the reconstructed $\pi^0\pi^0\ell\nu$ mode to the fit, and add one additional parameter that scales the non-resonant $\pi^+\pi^-\ell\nu$ and $\pi^0\pi^0\ell\nu$ rates.  We find we could project out a yield, relative to the charged $\rho$ yield, of $-2.2\% \pm 6.2\%$, consistent with zero.  

 \begin{sidewaystable}
 \begin{center}
 \caption{ \label{tab:expsyst}A summary of the experimental systematic uncertainties.  The bin numbers correspond to the $q^2$ and $\cos\theta_{Wl}$ intervals
that were independently reconstructed in the fit.
Errors from different sources are assumed to be uncorrelated.  Neutrino reconstruction systematic errors are the totals summarized in Table~\ref{tab:nureconsyst}. The phase space interval indices are as defined in Table~\ref{tab:intervals}.}
 \begin{tabular}{lcccccccccccccc}\hline\hline
  & \multicolumn{5}{c}{$\delta\mathcal{B}\left(B^0\to\pi^-\ell^+\nu\right)~[\%]$} & & \multicolumn{6}{c}{$\delta\mathcal{B}\left(B^0\to\rho^-\ell^+\nu\right)~[\%]$} & \multicolumn{2}{c}{$\delta\mathcal{B}\left(B^+\to(\eta\slash\eta^{\prime})\ell^+\nu\right)~[\%]$} \\
  \multicolumn{1}{c}{ {\bf Systematic Error Source} }& 1 & 2 & 3 & 4 & All & ~~~ & 1 & 2 & 3 & 4 & 5 & All & \hspace{25pt}$\eta$\hspace{25pt} & $\eta^{\prime}$ \\\hline
  Neutrino Reconstruction & 7.5 & 7.9 & 7.1 & 7.4 & 5.9 & & 21.1 & 26.2 & 16.0 & 11.6 & 71.5 & 8.9 & 23.4 & 19.3 \\
  Continuum Suppression & 9.8 & 3.0 & 0.9 & 1.1 & 1.1 & & 9.5 & 1.9 & 1.5 & 1.3 & 3.0 & 1.5 & 1.5 & 0.8 \\
  $B\to X_c\ell\nu$ & 4.0 & 3.9 & 1.3 & 1.2 & 1.5 & & 20.8 & 12.2 & 2.7 & 2.2 & 12.7 & 5.8 & 2.1 & 4.8 \\
  Other $B\to X_u\ell\nu$ & 2.3 & 1.2 & 2.3 & 5.8 & 2.5 & & 5.8 & 3.6 & 7.4 & 3.7 & 17.0 & 2.8 & 3.5 & 3.0 \\
  Fake Leptons & 4.9 & 1.3 & 2.3 & 0.6 & 1.7 & & 3.1 & 2.5 &  1.0 & 1.1 & 4.5 & 1.1 & 1.1 & 3.7 \\
  Lepton Identification & 2.0 & 2.0 & 2.0 & 2.0 & 2.0 & & 2.0 & 2.0 & 2.0 & 2.0 & 2.0 & 2.0 & 2.0 & 2.0 \\
  $\pi^0$ Identification & 0.3 & 0.3 & 0.2 & 0.2 & 0.1 & & 1.0 & 0.9 & 1.3 & 0.9 & 3.3 & 1.4 & 0.2 & 0.1 \\
  Number of $\Upsilon\to B\bar{B}$ & 3.6 & 3.6 & 3.6 & 3.6 & 3.6 & & 3.6 & 3.6 & 3.6 & 3.6 & 3.6 & 3.6 & 3.6 & 3.6\\
  $\tau_{B^+}/\tau_{B^0}$ & 0.5 & 0.3 & 0.4 & 0.4 & 0.4 & & 0.8 & 0.7 & 0.6 & 0.8 & 0.5 & 0.7 & 0.2 & 0.2 \\
  $f_{+-}/f_{00}$ & 0.5 & 0.9 & 0.7 & 0.8 & 0.7 & & 0.1 & 0.0 & 0.2 & 0.2 & 0.6 & 0.1 & 2.1 & 2.0 \\
  Non-Resonant $\pi\pi$ & 1.9 & 2.0 & 0.4 & 0.6 & 0.6 & & 5.1 & 3.6 & 2.1 & 3.2 & 3.9 & 3.5 & 2.0 & 2.1 \\
  Final State Radiation & 1.6 & 1.0  & 1.0 & 1.1 & 1.1 & & 1.2 & 1.5 & 2.2 & 1.8 & 2.0 & 1.8 & 1.2 & 1.6 \\ 
   \parbox[c]{1.5in}{ \vspace{0.1in} {\bf Sum } \vspace{0.1in} }  & {\bf 14.8} & {\bf 10.6 } & {\bf 9.1 } & {\bf 10.5} & {\bf 8.2} & & {\bf 32.5} & {\bf 29.8} & {\bf 18.7} & {\bf 13.7} & {\bf 75.3} & {\bf 12.6} & {\bf 24.4} & {\bf 21.2} \\
   \hline\hline
  \end{tabular}
  \end{center}
  \end{sidewaystable}
    
\subsection{Dependence on Form Factors}
 
 The previous CLEO measurement of  the $B\to\pi\ell\nu$ and $B\to\rho\ell\nu$ 
rates achieved a very minimal dependence on {\it a priori} form factors for
$B\to\pi\ell\nu$ by measuring independent rates in $q^2$ intervals similar to those in
this analysis.  Significant dependence on the $B\to\rho\ell\nu$ form factor, however, remained.
Adding the  $\cos\theta_{Wl}<0$ information in this analysis has greatly reduced
this residual dependence.
 
To evaluate the dependence on the form factors, we choose a set of 
calculations that span the a very conservative range of theoretical results.  For
each form factor, we re-weight our signal MC and refit to assess
the impact of the change.   We vary the $B\to\pi\ell\nu$ and $B\to\rho\ell\nu$ form factors
independently.   For each of those, we assign a systematic contribution for the rate in each measured phase space region  to be $1/2$ the difference of the largest and smallest rates obtained in that region for the different form factors.  The systematic errors due to uncertainties in the signal decay form factors are summarized in Table~\ref{tab:theosyst}. 

For  $B\to\pi\ell\nu$, we consider form factors from the unquenched lattice QCD calculations by HPQCD~\cite{HPQCD_2006} (our nominal form factor), from  Ball and Zwicky~\cite{ball_2005}, Scora and Isgur (ISGW2)~\cite{isgw2}, and 
Feldmann and Kroll~\cite{spd}.  The variation in $q^2$ dependence 
of the rate from these calculations is illustrated in 
Figure~\ref{fig:pi_formfactors}.  

For $B\to\rho\ell\nu$, we consider
the Ball and Zwicky LCSR calculation  (our nominal form factors), as well as those
of Melikhov and Stech~\cite{meli}, from  the quenched lattice calculations of UKQCD~\cite{ukqcd98}, and from Scora and Isgur (ISGW2)~\cite{isgw2}. 
Figure~\ref{fig:rho_formfactors} shows the the calculated rates as a function
 of $q^2$ and $\cos\theta_{Wl}$.
Note that the variation between calculations for the rate as a function of 
$\cos\theta_{Wl}$ is significant, which led to the remaining model dependence
of the previous analysis.

Our main sensitivity in the $B^+\to\eta^\prime\ell^+\nu$ mode comes from the $q^2<10$ GeV$^2$ region, increasing our sensitivity to this mode's form factor shape. We assess this sensitivity by varying the fraction of the $B\to\eta^\prime\ell\nu$ rate with $q^2<10$ GeV$^2$ by $\pm 10\%$ of itself.
 
\begin{table*}
\begin{center}
\caption{ \label{tab:theosyst}The systematic uncertainties associated with the individual form factor uncertainties are listed below.  The numbers correspond to the $q^2$ and $\cos\theta_{Wl}$ intervals
that were independently reconstructed in the fit.  The phase space interval indices are as defined in Table~\ref{tab:intervals}.}
  \begin{tabular}{lcccccccccccccc}\hline\hline
  & \multicolumn{5}{c}{$\delta\mathcal{B}\left(B^0\to\pi^-\ell^+\nu\right)~[\%]$} & & \multicolumn{6}{c}{$\delta\mathcal{B}\left(B^0\to\rho^-\ell^+\nu\right)~[\%]$} & \multicolumn{2}{c}{$\delta\mathcal{B}\left(B^+\to(\eta\slash\eta^{\prime})\ell^+\nu\right)~[\%]$} \\
  \multicolumn{1}{c}{ {\bf Form Factor} }& 1 & 2 & 3 & 4 & All & ~~~ & 1 & 2 & 3 & 4 & 5 & All & \hspace{25pt}$\eta$\hspace{25pt} & $\eta^{\prime}$ \\\hline
 $B\to\pi\ell\nu$ & 1.2 & 0.5 & 0.5 & 2.8 & 0.8 & & 0.3 & 0.3 & 0.7 & 0.3 & 2.1 & 0.4 & 0.3 & 0.2 \\
   $B\to\rho\ell\nu$ & 0.4 & 0.5 & 1.1 & 1.0 & 0.8 & & 2.9 & 4.9 & 0.9 & 1.8 & 7.0 & 1.8 & 0.3 & 1.6 \\
   $B\to\eta^\prime\ell\nu$ & 0.1 & 0.2 & 0.2 & 0.7 & 0.3 & & 0.9 & 1.5 & 1.1 & 0.9 & 5.2 & 0.5 & 0.3 & 1.4 \\
   \hline\hline
  \end{tabular}
  \end{center}
  \end{table*}

\section{Discussion of Results}
\label{sec:discussion}

\subsection{$|V_{ub}|$}

Extraction of $|V_{ub}|$ from a measured $B\to X_u\ell\nu$ rate requires theoretical input in the form of 
the form factor(s) governing their decay.   The unquenched LQCD calculations for the $B\to\pi\ell\nu$ form factor make a minimal number of assumptions.  Therefore, our primary result for $|V_{ub}|$ derives 
from our $B\to\pi\ell\nu$ measurement in combination with LQCD results.   The $\rho$ is unstable in unquenched lattice calculations, resulting in a multibody hadronic final state that current lattice technology is unable to accommodate.  Our primary result for $B\to\rho\ell\nu$ is a validation of the form factor shapes needed to control the $\rho\ell\nu$ backgrounds in other $B\to\pi\ell\nu$ measurements.

A determination of  $|V_{ub}|$ from our $B\to\pi\ell\nu$ results will be most competitive if the full range of
$q^2$ can be utilized.  However, unquenched LQCD calculations currently cover the range 
$q^2 > 16~\text{GeV}^2$~\cite{HPQCD_2006}.   The experimental constraints on the form factor shape do allow for useful extrapolation of the LQCD results outside this region~\cite{Arnesen:2005ez,Becher:2005bg}, but techniques for optimal use of both the experimental and LQCD information are still under development~\cite{Gibbons06}.  Unquenched lattice calculations that utilize a moving $B$ meson frame to extend the $q^2$ range~\cite{Davies:2006aa} are also under development, but not yet complete.
We therefore currently limit ourselves to the range $q^2 > 16~\text{GeV}^2$, for which the HPQCD collaboration finds $\Gamma(B^0\to\pi^-\ell^+\nu;\,q^2 > 16~\text{GeV}^2) = |V_{ub}|^2(2.07\pm 0.41 \pm 0.39 \text{~ps}^{-1})$~\cite{HPQCD_2006}.  Using a $B^0$ meson lifetime of $1.53\times 10^{-12}$~s~\cite{pdg06}, we find $|V_{ub}| = \left(3.6\pm 0.4 \pm 0.2^{+0.6}_{-0.4}\right)\times 10^{-3}$.  The errors are, in order,  statistical, experimental systematic, and theoretical systematic, where the final theoretical systematic error is driven by the expected uncertainty in the normalization of the form factor as calculated by our chosen LQCD calculation.  

\begin{figure}
\begin{center}
\includegraphics[width=3in]{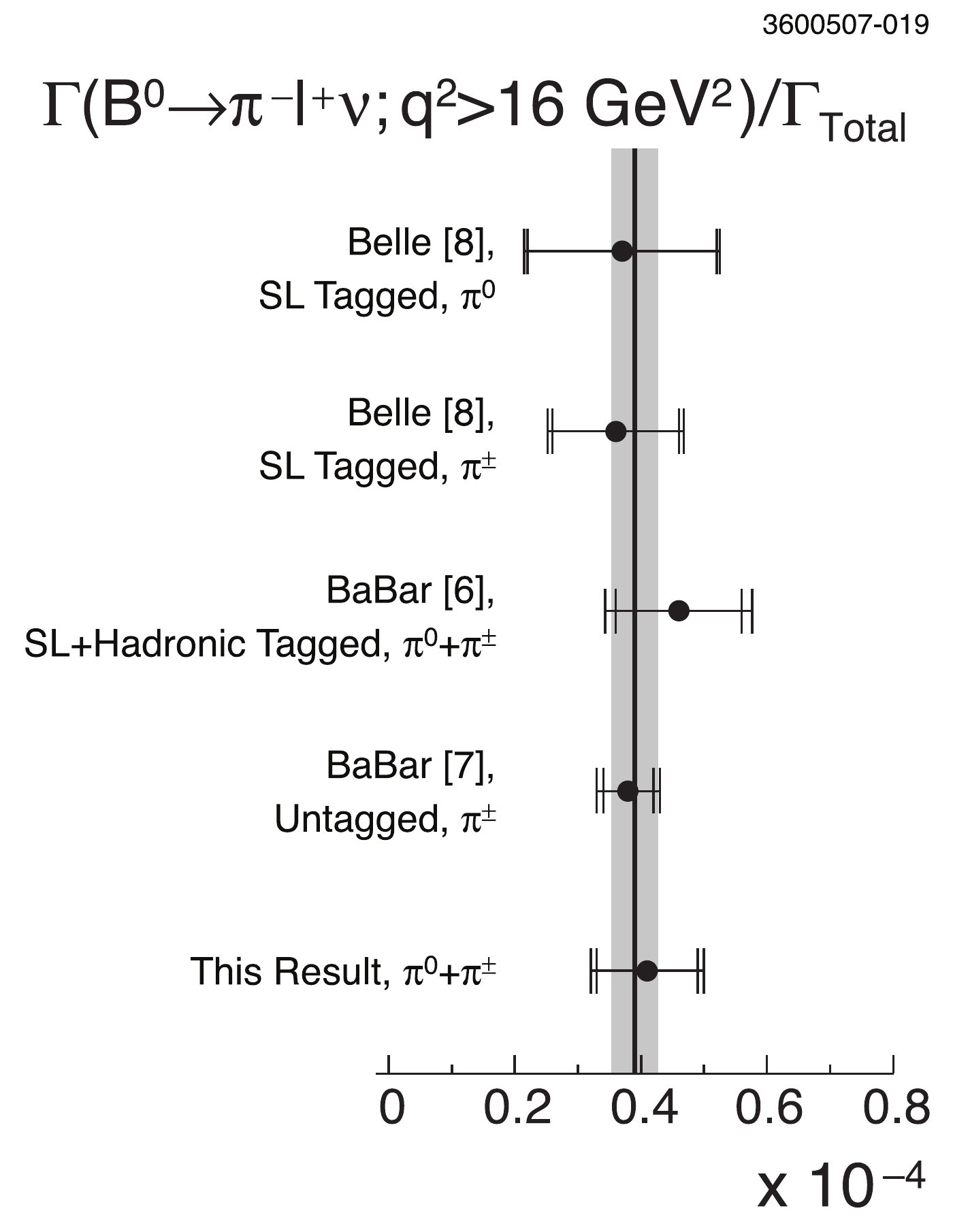}
\caption{\label{fig:pi_compare} A comparison of the current measurements of the $B^0\to\pi^-\ell^+\nu$ partial branching fraction in the $q^2 > 16$~GeV$^2$ region.  The inner (outer) error bars are statistical (statistical and systematic). Results from Belle~\cite{bb:BELLEy2007} use semleptonic $B$ decay to tag the non-signal $B$.  \textsc{BaBar} has published results for an analysis~\cite{bb:BABARy2006} using hadronic and semileptonic $B$ decays to tag the non-signal $B$ and also an analysis~\cite{bb:BABARy2007} utilizing loose $\nu$ reconstruction.  Where applicable $B^0\to\pi^-\ell^+\nu$ and $B^+\to\pi^0\ell^+\nu$ results have been averaged assuming isospin symmetry.  A weighted average and one standard-deviation error band, assuming uncorrelated errors, are shown. }
\end{center}
\end{figure}

Figure~\ref{fig:pi_compare} summarizes the current measurements by both Belle~\cite{bb:BELLEy2007} and \textsc{BaBar}~\cite{bb:BABARy2006,bb:BABARy2007} along with the results presented in this analysis for the $q^2>16~\text{GeV}^2$ region.  There very good agreement among the various results and analysis techniques for partial branching fraction in this region of phase space which is currently the key experimental input to determinations of $|V_{ub}|$ using unquenched LQCD calculations.  

\subsection{$B\to\rho\ell\nu$ Form Factors}
For $B\to\rho\ell\nu$, we perform a single-parameter fit to our measurements 
over the $\rho$ phase space for the  ISGW2~\cite{isgw2}, 
LCSR~\cite{ball_2005}, and quenched LQCD~\cite{DelDebbio:1997kr} (qLQCD)
 calculations.  The LCSR results are the results most heavily relied upon for
 simulation of the $\rho\ell\nu$ background to $\pi\ell\nu$.  The calculation 
is expected to be valid for the region  $q^2 \lesssim 14~\text{GeV}^2$.   
We therefore fit this calculation both to our measurements spanning the 
full $q^2-\cos\theta_{Wl}$ range and to our three subregions 
(see Table~\ref{tab:results}) that are restricted to $q^2 < 16~\text{GeV}^2$.  
The full phase space fits test the validity of a given model for use as a 
background model.  The restricted phase space fit for LCSR tests the shape 
without extrapolation outside its region of validity.  The results are 
summarized in Table~\ref{tab:rho_shape}.    All of the probabilities of
 $\chi^2$ exceed 10\% for the LCSR and LQCD, indicating reasonable agreement 
between theory and experiment at our current precision.  
The ISGW~II model has only a 5\% probability. 
While these constitute the most stringent tests to date of the $B\to\rho\ell\nu$
shape predictions, we cannot yet experimentally 
distinguish between the published central LCSR values and the effective shape variations used for
background systematics in the recent \textsc{BaBar} $B\to\pi\ell\nu$ measurement 
\cite{bb:BABARy2007}.   

\begin{table}
\caption{Summary of $\chi^2$ and number of degrees of freedom ($N_{\text{dof}}$) for fits to the $B\to\rho\ell\nu$ differential rate.}
\label{tab:rho_shape}
\begin{tabular}{lccc}\hline\hline
{\bf Form Factor} & ~~{\bf Phase Space}~~ & $\chi^2$ & $N_{\text{dof}}$  \\ \hline
ISGW~II & full           & 9.0 & 4  \\ 
LCSR  & full            & 4.5 & 4  \\
LCSR  & restricted & 4.3 & 2  \\ 
LQCD  & full            & 4.3 & 4 \\ \hline \hline 
\end{tabular}
\end{table}
  
\subsection{$\mathcal{B}(B^+\to \eta\ell^+\nu)$ and $\mathcal{B}(B^+\to\eta^{\prime}\ell^+\nu)$}
\label{sec:eta_discussion}

We do not see a statistically significant signal in the $B\to \eta\ell\nu$ mode.  When ${\cal B}(B^+\to \eta\ell^+\nu)$ is forced to zero, $-2\ln{\cal L}$ increases by 4.8, which corresponds to less than a three standard deviation effect.  While less statistically significant than reported by the previous analysis, the yields in the new relative to the old dataset are consistent within statistical errors.

Forcing ${\cal B}(B^+\to \eta^{\prime}\ell^+\nu)=0$, $-2\ln{\cal L}$ increases by 17.9, but we must fold in systematic uncertainties to evaluate the significance.   Taking the systematic uncertainties 
as Gaussian, we convolute them with our statistical probability distribution function (p.d.f.)
inferred from the $-2\ln{\cal L}$ distribution of the fit to model the total p.d.f..   From a toy MC based on this p.d.f., we determine the probability for measuring a rate greater than or equal to our central value, if the true rate were zero, to be $1.2\times10^{-3}$.  Therefore, we interpret the significance of our signal to be just over a three standard deviations.

Using the procedure just described, we find the 90\% confidence interval of 
$1.20\times 10^{-4}<{\cal B}(B^+\to\eta^{\prime}\ell^+\nu)<4.46\times 10^{-4}$.  For 
$B\to\eta\ell\nu$, we obtain a 90\% confidence level upper limit of ${\cal B}(B^+\to\eta\ell^+\nu) \le 1.01 \times 10^{-4}$.  We also determine the lower limit for the ratio
 ${\cal B}(B^+\to\eta^{\prime}\ell^+\nu)/{\cal B}(B^+\to\eta\ell^+\nu)$ to be 2.5 at the 90\% confidence level. 

Our results allow us to make some model dependent statements regarding the size of the extra gluon couplings from the 
$\eta^0$ singlet component of the $\eta$ and $\eta^{\prime}$. 
Beneke and Neubert use the Feldmann, Kroll, and Stech (FKS) mixing scheme~\cite{etap:FKS} in an analysis of $B\to\eta^{\prime}X_s$~\cite{etap:Neubert2003}
to model the $B\to\eta^{(\prime)}$ form 
factors as 
\begin{equation}
\label{EQEtapFF}
F^{B\to\eta^{(\prime)}}_+=F^{B\to\pi}_+\frac{f^q_{\eta^{(\prime)}}}{f_{\pi}} + F^{\text{singlet}}_+\frac{\sqrt{2}f^q_{\eta^{(\prime)}}+f^s_{\eta^{(\prime)}}}{\sqrt{3}f_{\pi}}, 
\end{equation}
where $F^{B\to\pi}$ is the  $B\to\pi$ form factor, $F^{\text{singlet}}$ is
an unknown singlet component,  and $f^{(q,s)}_{\eta^{(\prime)}}$ 
are constants determined by the FKS mixing scheme.
Kim~\cite{etap:CSK} recognized  that comparison of $B$ decay to $\eta\ell\nu$ and $\eta^{\prime}\ell\nu$
could determine the size of the singlet component.  

With the model-dependent parameter 
 \begin{equation}
\tilde{F}_s=\frac{\int |F^{\text{singlet}}_+|^2 \Omega_{\eta^{\prime}} \partial q^2}{\int |F^{B\to\pi}_+|^2 \Omega_{\eta^{\prime}} \partial q^2},
\end{equation}
where $\Omega_{\eta^{\prime}}$ is the appropriate phase space factor,
the $B^+\to\eta^{(\prime)}\ell^+\nu$ to $B^+\to\pi^0\ell^+\nu$ branching ratios may be expressed as
\begin{eqnarray}
\label{eq:etapBRexp}
\frac{{\cal B}_{\eta^{\prime}}}{2 {\cal B}_{\pi^0}} & = & \beta_{\eta^{\prime}} \left(  a_{\eta^{\prime}}^2 +2\gamma_{\eta^{\prime}}a_{\eta^{\prime}}b_{\eta^{\prime}}\sqrt{ \tilde{F}_s} + \tilde{F}_sb_{\eta^{\prime}}^2 \right) \\ \nonumber
\frac{{\cal B}_{\eta}}{2{\cal B}_{\pi^0}} & = & \beta_{\eta} \left(  a_{\eta}^2 +2\gamma_{\eta}a_{\eta}b_{\eta}\sqrt{ \tilde{F}_s \frac{\beta_{\eta^{\prime}}}{\beta_{\eta}}t} + \tilde{F}_s\frac{\beta_{\eta^{\prime}}}{\beta_{\eta}}tb_{\eta}^2 \right).
\end{eqnarray}
The parameters $a_{\eta^{(\prime)}}$ and  $b_{\eta^{(\prime)}}$ are combinations of
FKS mixing parameters, while $\beta_{\eta^{(\prime)}}$, 
$\gamma_{\eta^{(\prime)}}$ and $t$ capture differences in the $F^{\pi}_+$ and $F^{\text{singlet}}_+$ 
$q^2$ dependence.  Following Kim,  we assume that the $F^{B\to\pi}$ and $F^{\text{singlet}}$ 
have the same $q^2$ dependence, and cover shape differences with a systematic uncertainty.

\begin{figure}
\begin{center}
\includegraphics[width=3in]{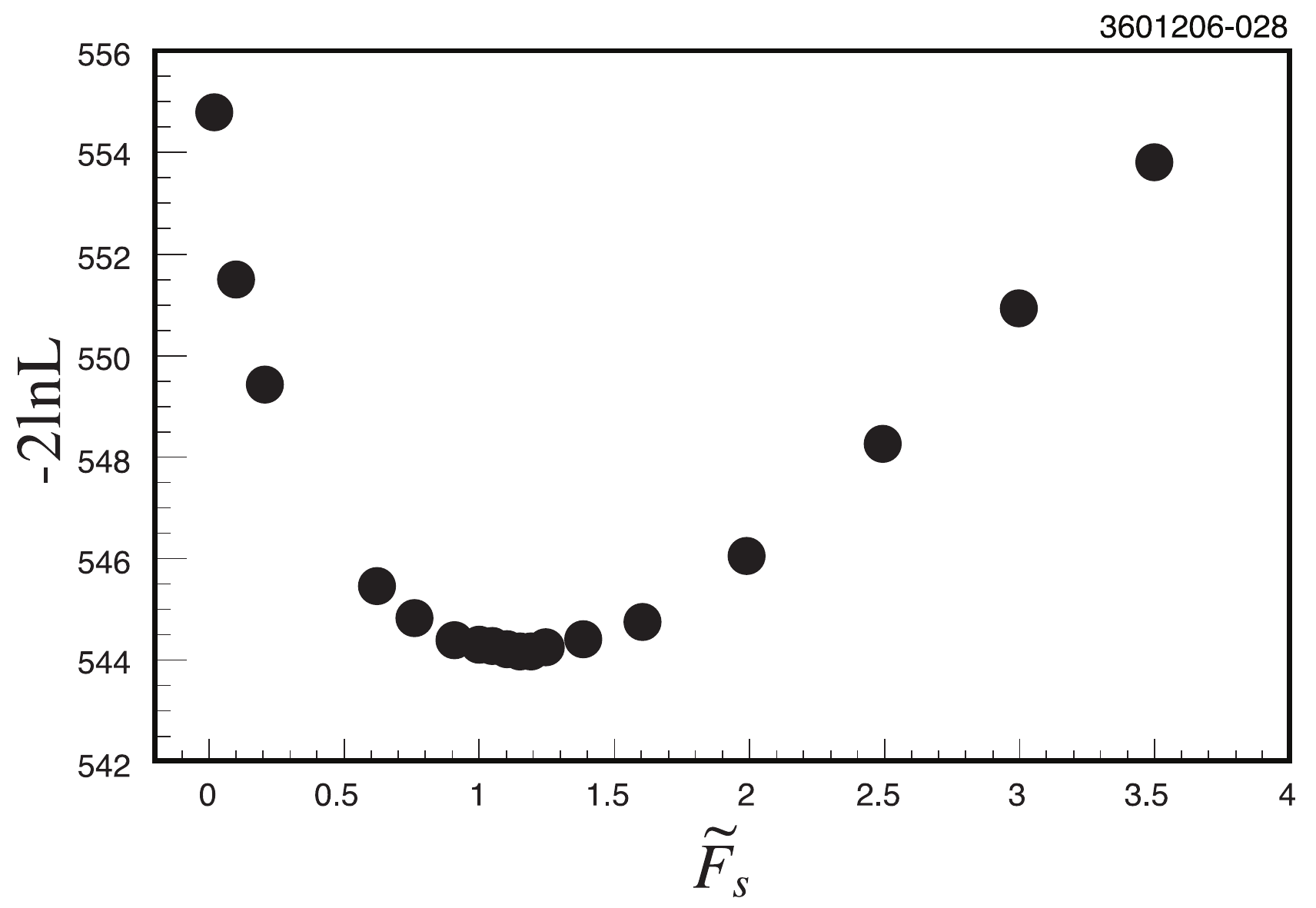}
\caption{\label{fig:fs_ll_dist} The change in the $-2\ln{\cal L}$ of the fit as a function of the parameter $\tilde{F}_s$.}
\end{center}
\end{figure}

We impose  these branching ratio relationships in our standard fitting procedure and allowing $\tilde{F}_s$ to float as a free parameter.  We find
$\tilde{F}_s =  1.15\pm0.54\pm0.38\pm0.21$, 
where the errors listed are, in order, statistical, systematic, and those due to uncertainties related to this model, including the relative $q^2$ dependence.  The variation of the likelihood with $\tilde{F}_s$ is shown in Figure~\ref{fig:fs_ll_dist}. The $B\to\eta^{(\prime)}\ell\nu$
branching fractions found with this model are consistent with the results of our nominal fit. 

With our $\tilde{F}_s$ determination, we can roughly update the 
$B^-\to K^-\eta^{\prime}$ prediction of Benke and Neubert by
scaling their $F^{\text{singlet}}(q^2=0)=0.1$ 
(about equivalent to  $\tilde{F}_s \sim 0.137$) result,
 for which
they found ${\cal B}(B^-\to K^-\eta^{\prime})=(56^{+19}_{-14}{}^{+31}_{-13})\times 10^{-6}$.
Experimentally, ${\cal B}(B^-\to K^-\eta^{\prime})=(71\pm4)\times 10^{-6}$~\cite{pdg06}.   With
our $\tilde{F}_s$, we estimate ${\cal B}(B^-\to K^-\eta^{\prime})=(84^{+35}_{-25}{}^{+53}_{-24})\times 10^{-6}$, though a
rigorous calculation with $\tilde{F}_s \sim 1$ is needed.  This estimate is double the 
prediction with no singlet contribution and  agrees well with the 
experimental value. 

\section{Summary}

We have measured the branching fractions $\mathcal{B}\left(B^0\to\pi^-\ell^+\nu\right)=(1.37 \pm 0.15 \pm 0.11)\times 10^{-4}$ and $\mathcal{B}\left(B^0\to\rho^-\ell^+\nu\right) = (2.93 \pm 0.37 \pm 0.37)\times 10^{-4}$ with very little residual dependence of either branching fraction on either the $B\to \pi$ or $B\to \rho$ form factors.  Table~\ref{tab:results} summarizes the branching fraction results for the partial phase space measurements that are summed to obtain the measurements integrated over phase space.
These results agree well with recent measurements from \textsc{BaBar}~\cite{bb:BABARy2005,bb:BABARy2006,bb:BABARy2007}
and Belle~\cite{bb:BELLEy2007}.  The total branching fractions for the $\pi\ell\nu$ and $\rho\ell\nu$ modes are among the most precise current measurements.  These results indicate that at the level of experimental precision we can probe, the $B\to\rho\ell\nu$ form factor shapes obtained with LCSR agree with our data, though  we cannot test the shape at the uncertainty level assumed in the \textsc{BaBar} $B\to\pi\ell\nu$ analysis~\cite{bb:BABARy2007}. Using the most recent unqenched lattice QCD calculations of the $B\to\pi\ell\nu$ form factor~\cite{HPQCD_2006}, and measured rate for  $q^2>16~\text{GeV}^2$ region we extract $|V_{ub}| = (3.6 \pm 0.4 \pm 0.2 ^{+0.6} _{-0.4})\times 10^{-3}$, where the uncertainties are statistical, experimental, and theoretical, respectively. While this value is competitive with other recent determinations of $|V_{ub}|$, the full strength of this analysis will be more effectively realized when a broader $q^2$ range can be used reliably for the determination of $|V_{ub}|$.

We find evidence for $B^+\to\eta^{\prime}\ell^+\nu$  at the three standard deviation 
level with a  branching fraction of
${\cal B}(B^+\to\eta^{\prime}\ell^+\nu) = (2.66 \pm 0.80 \pm  0.56 ) \times10^{-4}$ and a 90\% confidence interval of
$1.20\times 10^{-4}<{\cal B}(B^+\to\eta^{\prime}\ell^+\nu)<4.46\times 10^{-4}$.  The probability that our results for $\mathcal{B}\left(B^+\to\eta^{\prime}\ell^+\nu\right)$ are consistent with a previous 90\% confident limit set by \textsc{BaBar}~\cite{BaBar_eta_ichep06} of $1.3\times 10^{-4}$ is approximately 5\%.
We establish an  upper limit for the $B^+ \to\eta\ell^+\nu$ branching fraction of
${\cal B}(B^+\to\eta\ell^+\nu) \le 1.01 \times 10^{-4}$ (90\% confidence level), with systematic uncertainties included.  This is consistent with the  \textsc{BaBar} 
 90\% confident upper limit~\cite{BaBar_eta_ichep06} of $1.4 \times 10^{-4}$. These results
 imply  a  lower limit of ${\cal B}(B^+\to\eta^{\prime}\ell^+\nu)/{\cal B}(B^+\to\eta\ell^+\nu) >2.5$ (90\% confidence level).  Furthermore, the relative rates indicate a significant form factor
contribution from the singlet component of the  $\eta^{\prime}$, which can
bring predictions for ${\cal B}(B\to K^-\eta^{\prime})$ into better agreement
with experimental measurements.

These results supersede those previously published by CLEO~\cite{bb:Athar:2003yg}.  
While some shifts of central values are observed, particularly for $B^0\to\rho^-\ell\nu$, these results are consistent within statistical and systematic errors of those previously presented.  We note that retuning of the event selection algorithm at low $q^2$ significantly reduces the statistical and systematic correlation between results in this region of phase space.  Furthermore, expanding the measured $B\to\rho\ell\nu$ phase space nearly eliminates the theoretical systematic errors due to the $B\to\rho$ form factor uncertainty that were a significant contribution to the total error on the previous result.

To allow external use of our data in form factor,  $|V_{ub}|$, and $\eta$ and $\eta^{\prime}$ isosinglet studies, we provide the full statistical and systematic uncertainty correlation matrices in Appendix~A.

We gratefully acknowledge the effort of the CESR staff in providing us with excellent luminosity and running conditions. D.~Cronin-Hennessy and A.~Ryd thank the A.P.~Sloan Foundation. This work was supported by the National Science Foundation, the U.S. Department of Energy, and the Natural Sciences and Engineering Research Council of Canada.

\section{Appendix:  Full Correlation Matrices}

Tables~\ref{tab:stat_corr_matrix} and~\ref{tab:syst_corr_matrix} provide the correlation matrices for both the statistical and systematic errors on the branching fractions outlined in Table~\ref{tab:results}.  Cross-feed among bins produces anti-correlations in the results for individual bins; therefore, techniques for extracting $|V_{ub}|$ by averaging over multiple bins in the fit tend to have reduced uncertainty when these correlations are taken into account.

\begin{table*}
\begin{center}
\caption{ \label{tab:stat_corr_matrix} The statistical correlation matrix for the measured branching fractions. The phase space interval indices are as defined in Table~\ref{tab:intervals}.}
\begin{tabular}{ccccccccccccc} \hline\hline
 {\bf Mode} & {\bf Index} \\  \hline
 $\pi$ & 1  & 1.000 & -0.098 & 0.001 & -0.003 & -0.002 & 0.004 & -0.158 & 0.036 & 0.008 & 0.007 & 0.003 \\ 
 $\pi$ & 2  & -0.098 & 1.000 & -0.040 & 0.001 & -0.007 & 0.012 & -0.015 & -0.195 & 0.037 & -0.003 & 0.007 \\ 
 $\pi$ & 3 & 0.001 & -0.040 & 1.000 & -0.020 & -0.025 & 0.013 & 0.003 & -0.071 & -0.173 & 0.089 & -0.105 \\ 
 $\pi$ & 4 & -0.003 & 0.001 & -0.020 & 1.000 & -0.054 & 0.018 & -0.013 & -0.019 & -0.015 & -0.345 & -0.172 \\ 
 $\eta$ & -- & -0.002 & -0.007 & -0.027 & -0.054 & 1.000 & -0.008 & -0.001 & 0.002 & 0.012 & -0.058 & -0.059 \\ 
 $\eta^{\prime}$ & --  & 0.004 & 0.012 & 0.013 & 0.018 & -0.008 & 1.000 & -0.010 & -0.054 & -0.043 & 0.004 & -0.113 \\ 
 $\rho$ & 1 & -0.158 & -0.015 & 0.003 & -0.013 & -0.001 & -0.010 & 1.000 & -0.151 & 0.031 & -0.007 & 0.048 \\ 
 $\rho$ & 2  & 0.036 & -0.195 & -0.071 & -0.019 & 0.002 & -0.054 & -0.151 & 1.000 & 0.027 & 0.063 & 0.102 \\ 
 $\rho$ & 3  & 0.008 & 0.037 & -0.173 & -0.015 & 0.012 & -0.043 & 0.031 & 0.027 & 1.000 & -0.126 & 0.036 \\ 
 $\rho$ & 4  & 0.007 & -0.003 & 0.089 & -0.345 & -0.058 & 0.004 & -0.007 & 0.063 & -0.126 & 1.000 & -0.267 \\ 
 $\rho$ & 5  & 0.003 & 0.007 & -0.105 & -0.172 & -0.059 & -0.113 & 0.048 & 0.102 & 0.036 & -0.267 & 1.000 \\ 
 \hline\hline
\end{tabular}
\end{center}
\end{table*}
 
\begin{table*}
\begin{center}
\caption{ \label{tab:syst_corr_matrix} The systematic correlation matrix for the measured branching fractions.  The phase space interval indices are as defined in Table~\ref{tab:intervals}.}
\begin{tabular}{ccccccccccccc} \hline\hline
 {\bf Mode} & {\bf Index} \\  \hline
 $\pi$ & 1 & 1.000 & 0.674 & 0.678 & 0.383 & 0.388 & -0.033 & 0.501 & 0.373 & 0.337 & 0.406 & -0.257  \\ 
 $\pi$ & 2 & 0.674 & 1.000 & 0.754 & 0.115 & 0.575 & -0.000 & 0.279 & 0.012 & 0.604 & 0.661 & 0.246  \\ 
 $\pi$ & 3 & 0.678 & 0.754 & 1.000 & 0.588 & 0.502 & -0.016 & 0.368 & 0.258 & 0.536 & 0.514 & 0.056  \\ 
 $\pi$ & 4 & 0.383 & 0.115 & 0.588 & 1.000 & 0.407 & -0.340 & 0.563 & 0.704 & 0.111 & -0.099 & -0.542  \\ 
$\eta$ & -- & 0.388 & 0.575 & 0.502 & 0.407 & 1.000 & -0.665 & 0.416 & 0.406 & 0.587 & 0.260 & -0.299  \\ 
$\eta^{\prime}$ & -- & -0.033 & -0.000 & -0.016 & -0.340 & -0.665 & 1.000 & -0.285 & -0.477 & -0.232 & 0.228 & 0.571  \\ 
$\rho$ & 1 & 0.501 & 0.279 & 0.368 & 0.563 & 0.416 & -0.285 & 1.000 & 0.852 & 0.200 & 0.110 & -0.580  \\ 
$\rho$ & 2 & 0.373 & 0.012 & 0.258 & 0.704 & 0.406 & -0.477 & 0.852 & 1.000 & 0.136 & -0.159 & -0.831  \\ 
$\rho$ & 3 & 0.337 & 0.604 & 0.536 & 0.111 & 0.587 & -0.232 & 0.200 & 0.136 & 1.000 & 0.737 & 0.141  \\ 
$\rho$ & 4 & 0.406 & 0.661 & 0.514 & -0.099 & 0.260 & 0.228 & 0.110 & -0.159 & 0.737 & 1.000 & 0.442  \\ 
$\rho$ & 5 & -0.257 & 0.246 & 0.056 & -0.542 & -0.299 & 0.571 & -0.580 & -0.831 & 0.141 & 0.442 & 1.000  \\ \hline\hline
\end{tabular}
\end{center}
\end{table*}

\end{document}